\documentclass[a4paper,
preprint,
aps,pra,superscriptaddress,floatfix,showpacs,showkeys]{revtex4-1}

\usepackage[english]{babel}
\usepackage{amsmath}
\usepackage{amssymb,graphicx}
\usepackage{braket}
\usepackage{ifpdf,color,tabularx,multirow}

\usepackage{color}
\definecolor{darkblue}{rgb}{0,0,.6}
\definecolor{darkgreen}{rgb}{0,0.5,0}

\usepackage{ifthen}

\definecolor{darkblue}{rgb}{0,0,.6}

\ifpdf
  \usepackage{epstopdf}
  \usepackage[pdftex,unicode,
  pdfstartview={FitH},pdfborder={0 0 0}]{hyperref}
  \usepackage{hypcap}
\else
  \usepackage[hypertex]{hyperref}
\fi
\hypersetup{
  bookmarksnumbered = true,
  colorlinks = true, linkcolor = darkblue,
  citecolor = darkblue, filecolor = darkblue,
  menucolor = darkblue, urlcolor = darkblue
}

\usepackage{bm}

\let\vec=\bm

\newcolumntype{C}[1]{>{\centering\arraybackslash}m{#1}}

\DeclareMathOperator{\Tr}{Tr}

\renewcommand\Re{\operatorname{Re}}

\newcommand{\abs}[1]{\left|#1\right|}

\newcommand{\ee}{\mathrm{e}}

\newcommand{\Eex}{\ensuremath{E_{\mathrm{ex}}}}
\newcommand{\ELO}{\ensuremath{\epsilon_{\mathrm{LO}}}}
\newcommand{\Eg} {\ensuremath{E_\mathrm{g}}}

\newcommand{\kB} {\ensuremath{k_\mathrm{B}}}

\newcommand{\ta}[1]{\tau_{\mathrm{#1}}}

\renewcommand{\H}[1]{H_\mathrm{#1}}
\newcommand{\rh}[1]{\rho_\mathrm{#1}}
\newcommand{\w}[1]{\omega_\mathrm{#1}}

\newcommand{\trh}[1]{\widetilde\rho_\mathrm{#1}}

\renewcommand{\H}[1]{H_\mathrm{#1}}
\newcommand{\US}{U_\mathrm{S}}
\newcommand{\aSiO}{\ensuremath{\alpha\text{-SiO}_2}}

\begin{document}
\title{Non-Markovian pure dephasing in a dielectric excited by a few-cycle laser pulse}
\author{Stanislav Yu.~Kruchinin}
\email{stanislav.kruchinin@univie.ac.at}

\affiliation{Center for Computational Materials Sciences, Faculty of Physics, University of Vienna, Sensengasse 8/12, 1090 Vienna, Austria}

\begin{abstract}
We develop the theory of pure dephasing in a solid exposed to an ultrashort laser pulse beyond the commonly used Markov approximation.
This approach takes into account the finite cutoff energy of the bath and can be applied to both many-particle and phonon environments.
With numerical simulations performed with the time-dependent Hartree-Fock equations, we investigate how the excitation probability and high-harmonic generation are described by different models of decoherence.
It is shown that the time-dependent rates allow for temporally high dephasing to successfully reproduce the main features of high-harmonics spectrum and avoid an overestimation of the charge carrier population after the pulse, which is a common problem of the Markov approximation.
\end{abstract}

\pacs{78.20.Bh, 72.80.Sk, 77.22.Ej, 42.50.Hz}

\keywords{%
decoherence, dielectric, semiconductor, few-cycle pulse, strong field, ultrafast optics
}
\maketitle

\section{Introduction}

Recent progress in the synthesis of laser waveforms in the IR and visible domains~\cite{Goulielmakis_2008_SCI_320_1614, Huang_2011_NP_5_475, Fattahi_2014_Optica_1_45} have stimulated experimental and theoretical investigations of strong-field phenomena and quantum control in wide bandgap insulators~\cite{Schiffrin_2013_Nature_493_70, Luu_2015_Nature_521_498}, two-dimensional~\cite{Liu_2017_NP_13_262, Kelardeh_2016_PRB_93_155434, Motlagh_2018_PRB_98_125410} and nanostructured materials~\cite{Han_2016_NC_7_13105, Vampa_2017_NP_13_659}.

The majority of modern theoretical treatments of these systems are employing the Markov approximation describing the relaxation phenomena by constant rates $1/T_1$ and $1/T_2$.
This approach was successfully applied in the recent studies of high harmonic generation (HHG) in solids~\cite{Ghimire_2011_NP_7_138, Vampa_2014_PRL_113_073901, Luu_2015_Nature_521_498, Garg_2016_Nature_538_7625, Ndabashimiye_2016_Nature_534_520} provided experimental methods allowing for a distinction of the inter- and intraband components of polarization via analysis of the spectrum, waveform and group delay of the emitted radiation.
Notably, the HHG measurements in thin films of fused silica have shown that the group delay and scaling of the cutoff in high-frequency plateaus with the field amplitude demonstrate the features of an intraband current~\cite{Garg_2016_Nature_538_7625, Garg_2018_NP_12_291}, whereas the measurements in a semiconductor (GaSe) demonstrate the leading role of interband transitions and their interference~\cite{Schubert_2014_NP_8_119, Hohenleutner_2015_Nature_523_572}.

The state-of-the-art quantum-mechanical models significantly overestimate the interband polarization and require very short dephasing time $T_2 \approx 0.4 - 4$~fs for reproducing the experimental data in SiO$_2$~\cite{Garg_2016_Nature_538_7625, Garg_2018_NP_12_291} and ZnO~\cite{Ghimire_2011_NP_7_138, Vampa_2014_PRL_113_073901}.
This assumption is incompatible with simulations of carrier-envelope-phase (CEP) control of the current in dielectrics and semiconductors~\cite{Schiffrin_2013_Nature_493_70, Kruchinin_2013_PRB_87_115201, Paasch-Colberg_2016_Optica_3_1358} and the dynamic Franz-Keldysh effect~\cite{Schultze_2013_Nature_493_75, Schultze_2014_Sci_346_1348, Lucchini_2016_Science_353_916}, where ultrafast dephasing was not required for reaching an agreement with experiment.

The recent multiscale ab initio simulations of HHG in a diamond~\cite{Floss_2018_PRA_97_011401}, considered additional averaging by intensity distribution in the laser beam spot and emphasized the role of propagation effects in the build-up of a smooth harmonic spectrum.
Nevertheless, these additional considerations still cannot reproduce the absence of a group delay dispersion in the emitted radiation~\cite{Garg_2016_Nature_538_7625, Garg_2018_NP_12_291}, which is peculiar to the intraband current, and simulations with insufficiently high decoherence rates still predict the dominant contribution of interband polarization in the high-frequency plateaus of the spectrum~\cite{Otobe_2016_PRB_94_235152}.
On one hand, assumption of ultrafast scattering times $\sim 0.1 - 4$~fs previously reported in the semiclassical simulations of a high-field transport in SiO$_2$~\cite{Fischetti_1985_PRB_31_8124, Arnold_1994_PRB_49_10278} and fully microscopic simulations of nonlinear spectroscopy in semiconductors~\cite{Kira_2006_PQE_30_155, Smith_2010_PRL_104_247401} increases the intraband component of the current density to the reproduces both the high-energy plateaus of experimental HHG spectra and group delay~\cite{Vampa_2015_PRL_115_193603, Luu_2015_Nature_521_498, Garg_2016_Nature_538_7625, Garg_2018_NP_12_291}.
On the other hand, as will be shown below, an ultrafast dephasing rate results in the overestimated spectral broadening and wrong scaling of the carrier population with the field intensity due to the opening of an artificial single-photon excitation channel.
The recently reported experimental results on the optically-controlled current suggest that the total charge induced by the laser pulse in a circuit scales close to the perturbative result at low intensities scales when the system is still in the multiphoton regime~\cite{Schiffrin_2013_Nature_493_70, Paasch-Colberg_2016_Optica_3_1358}.
Thus, a more appropriate model of dephasing in dielectrics needs to be developed.

In this work, we theoretically investigate pure dephasing in a dielectric beyond the Markov approximation~\cite{Tokuyama_1976_PTP_55_411, Ahn_1994_PRB_50_8310, Breuer_2016_RMP_88_021002, deVega_2017_RMP_89_015001}.
We show that it is possible to introduce the dephasing rate as a time-dependent function with a slowly-varying envelope accounting for a non-trivial bath spectral function.
Starting from the model Hamiltonian, we derive the semiconductor Bloch equations applicable for a description of pure dephasing beyond the Markov approximation.
Both environments are analyzed in the framework of the harmonic oscillator model, where finite spectral cutoff energy is taken into account.
The strong dependence of dephasing rate on the cutoff energy explains why the phonon bath is approximately Markovian even on a few-femtosecond time scale and why the many-particle environment features a temporally high dephasing rate, which has not been observed in the Markovian limit.
These results are illustrated by numerical simulations in both independent-particle and time-dependent Hartree-Fock approximations.
It is shown that the time-dependent rate given by the excitation-induced dephasing with the envelope of harmonic oscillator improves the field amplitude scaling of excitation probability and produces high-harmonic spectra in a good agreement with experimental data.

\section{Approximations and applicability limits}

\begin{figure}[!htb]
  \includegraphics[width=\linewidth]{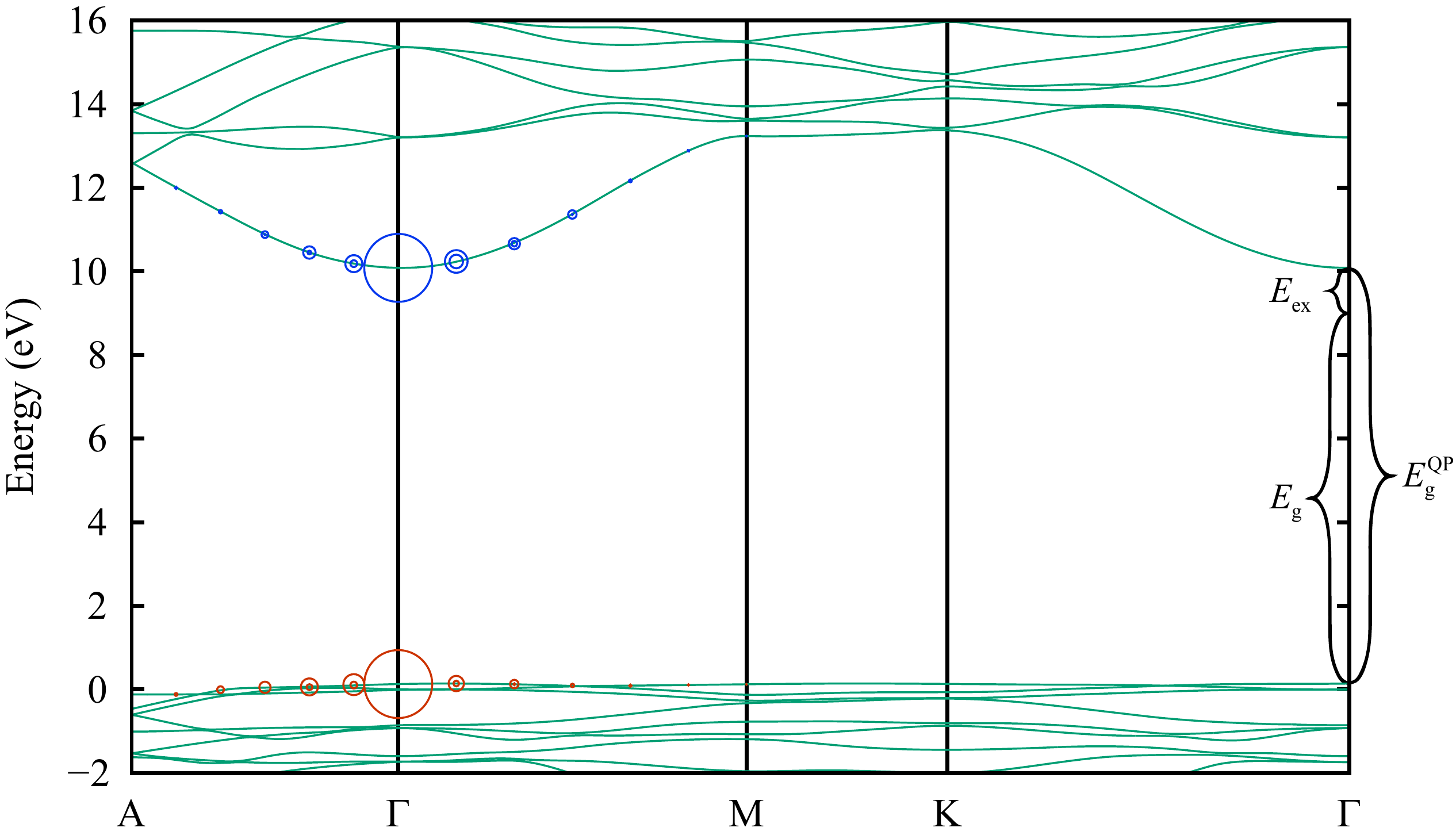}
  \caption{\label{f:bands}%
    Band structure of \aSiO{} in the self-consistent quasiparticle $GW$ approximation (solid lines) and the fat band plot (circles) calculated from the Bethe-Salpeter equation (BSE) implemented in the VASP code~\cite{Kresse_2012_PRB_85_045205, Sander_2015_PRB_92_045209}.
    Circle size is proportional to the squared modulus of the exciton wavefunction expansion coefficient and show which electron-hole pairs contribute the $1S$ excitonic state.
    For this simulation we used the $\Gamma$-centered $k$ grid with $13 \times 13\times 13$ $\vec k$ points and interpolated energy bands with the \textsc{wannier90} program~\cite{Mostofi_2014_CPC_185_2309}.
 }
\end{figure}

We start from a discussion of material parameters in dielectrics and semiconductors to validate the standard approximations applied in simulations of ultrafast spectroscopy.
Fig.~\ref{f:bands} shows the state-of-the-art \emph{ab initio} simulation of \aSiO{} band structure in the self-consistent quasiparticle $GW$ approximation.
The circles show relative contributions of the Bloch orbitals to the excitonic $1S$-state determined from the solution of the Bethe--Salpeter equations.
The direct quasiparticle bandgap at the $\Gamma$ point is $\Eg^\text{QP} = 10.1$~eV, and the optical bandgap $\Eg = 8.9$~eV is shifted from it by the exciton binding energy $\Eex = 1.2$~eV obtained by solving of the Bethe-Salpeter equations.
Remarkably, the exciton binding energy in \aSiO{} is by 2--3 orders of magnitude larger than that in the commonly studied semiconductors, e.g., $\Eex^\mathrm{GaAs} = 4$~meV~\cite{Nam_1976_PRB_13_761}, $\Eex^\mathrm{Si} = 15$~meV~\cite{Green_2013_AIPA_3_112104}.
The value of $\Eex = 1.2$~eV is larger than the Rabi frequency $\omega_\mathrm{R}^{\mathrm{max}} = F_0 \max_{\vec k} |\vec \xi_{nm}(\vec k)| \approx 0.74$~eV even at the field amplitude of $F_0 = 2$~V/\AA{} close to the damage threshold, where $\max_{\vec k} |\vec \xi_{nm,\vec k}| \approx 0.37$~\AA{}.~\cite{Kruchinin_2018_RMP_90_021002}
Therefore, electron-electron interaction plays a significant role in electron dynamics of the wide bandgap dielectrics, even though the excitonic peaks are broadened and not visible in the absorption spectrum at high intensities.
This statement is also true for some other wide bandgap materials, e.g., CaF$_2$~\cite{Tsujibayashi_2002_REDS_157_969, Sugiura_2016_JJAP_31_2816}, where the exciton binding energy is also $\sim 1$~eV.

\begin{table}[!htb]
  \caption{\label{t:times}%
    Characteristic time scales of the field-matter interaction in the Houston basis and their typical values for phonon (ph) and many-particle environments (mp).
    The parameters are estimated for \aSiO{} ($\Eg = 8.9$~eV, $\Eex = 1.2$~eV, $\ELO = 150$~meV) and VIS/NIR laser pulses in the wavelength range $400 - 2500$~nm with the field amplitude $F_0 = 1$~V/\AA{}.
  }
  \begin{tabular*}{\linewidth}{m{0.38\linewidth}m{0.3\linewidth}m{0.32\linewidth}}
    \hline\hline
    Time scale & Denotation & Values\\\hline
    Optical cycle  & $T_0 = 2\pi/\omega_0$ & $1.4-8.4$~fs\\
    Pulse duration & $\ta{L}$ & $0.7-40$~fs\\
    Elapsed time   & $t_\mathrm{max}$  & $\sim \ta{L}$\\
    Minimal bandgap
    & $\ta{g} = 2\pi/(\Eg + U_\mathrm{p})$
    & $0.23 - 0.46$~fs\\
    Change of adiabatic energies
    & $\ta{AE}$
    & $\gtrsim 0.1$~fs\\
    Change of adiabatic states
    & $\ta{AS}$
    & $1-8$~fs\\
    Minimal relaxation time
    & $\ta{relax.}$
    & $\gtrsim 27.6$~fs (ph), $\sim 3.5$~fs (mp)\\
    Bath correlation decay time
    & $\ta{corr.}$
    & $\sim 10$~fs (ph), $\sim 2$~fs (mp)\\
    \hline\hline 
  \end{tabular*}
\end{table}

\begin{table}
  \caption{\label{t:approx}%
    Summary of approximations applicability for the bath of phonons and many-body correlations.
  }
  \begin{tabular*}{\linewidth}{m{0.35\linewidth}m{0.35\linewidth}C{0.15\linewidth}C{0.15\linewidth}}
    \hline\hline
    \multirow{2}{*}{Approximation} & \multirow{2}{*}{Condition} & \multicolumn{2}{c}{Applicability} \\
    &  & ph & mp \\
    \hline
    Weak coupling (Born) & $\ta{corr.} \ll \ta{relax.}$   & Yes & Yes\\
    Secular              & $\ta{g} \ll \ta{relax.}$; $\ta{g} \ll \ta{AE}, \ta{AS}$  & Yes & Partial\\
    Instantaneous eigenbasis & $\ta{corr.} \ll \ta{AE}, \ta{AS}$ & Yes & Partial \\
    Markov               & $\ta{corr.} \ll t_\mathrm{max}$ & Yes  & No \\
    \hline\hline
  \end{tabular*}
\end{table}

For convenience, we summarized the characteristic time scales of the laser-matter interaction problem for a representative material ($\alpha$-quartz) and applicability conditions of the relevant approximations in the Tables~\ref{t:times} and \ref{t:approx}, respectively.
The temporal change of adiabatic eigenenergies and eigenstates are described by the parameters
$\ta{AE} \equiv 2\pi/\max \abs{\frac{1}{E_{n\vec K(t)}} \frac{d}{dt} E_{n\vec K(t)}}$ and
$\ta{AS} \equiv 2\pi/\max\abs{\langle u_{n\vec K(t)} | \frac{d}{dt} | u_{m\vec K(t)}\rangle} 
= 2\pi/\max_{\vec k} |\vec F_0 \cdot \vec\xi_{nm,\vec K(t)}|$, respectively,
$\vec K(t) = \vec k + \vec A(t)$ is the time-dependent crystal momentum given by the acceleration theorem~\cite{Bloch_1928_ZP_52_555}, $\vec A(t) = - \int_0^t dt'\vec F(t')$ is the vector potential, and $\vec\xi_{nm,\vec k} = \braket{ u_{n\vec k} | i\partial_{\vec k} | u_{m\vec k}}$ is the matrix element of a coordinate operator in the crystal momentum representation.
Our ab initio simulations show that in \aSiO{} and other dielectrics with large effective masses of carriers, the matrix elements $\vec\xi_{nm, \vec k}$ are slowly-varying functions of $\vec k$, and thus $\ta{AE} \ll \ta{AS}$.
In the materials with small effective masses of electrons in the conduction band, e.g., GaAs,~\cite{Wismer_2016_PRL_116_197401} the optical matrix element change more rapidly with $\vec k$ than the band energies, so the opposite situation ($\ta{AE} \gg \ta{AS}$) might be realized as well.

\section{Non-Markovian master equations}

If the system's evolution time is much longer than the correlation decay time, the scattering events can be viewed as instantaneous in comparison to evolution time, which leads to the Markov approximation with constant dephasing rates $\propto 1/T_2$.
As shown in the previous section (see Table~\ref{t:approx}), the many-particle environment does not satisfy this condition, since the correlation decay time is comparable to the evolution time $\ta{corr.} \gtrsim t_\mathrm{max}$.
The fat band plot in Fig.~\ref{f:bands} shows that the $1S$ excitonic state is formed primarily by the Bloch orbitals at $\vec k = 0$ and suggests that it is possible to adiabatically separate the Hilbert space.
The fast-evolving single-particle states participate in the intraband motion and dynamic Bloch oscillations, while the slow carriers stay in the middle of BZ and form the many-particle states.

The full Hamiltonian is written as
\begin{equation}
  H(t) = \H{S} + \H{B} + \H{SB}(t),
\end{equation}
where $\H{S}(t) \equiv H_0 + \H{L}(t)$,
\begin{equation*}
  H_0 = \sum_{n,\vec k} E_{n\vec k} a_{n\vec k}^\dagger a_{n\vec k}
\end{equation*}
is the Hamiltonian of the single-particle states occupying the accelerated Bloch (Houston) state at the band $n$ and initial crystal momentum $\vec k$,
$a_{n\vec k}^{\phantom\dagger}$ and $a_{n\vec k}^\dagger$ are their ladder operators,
$\H{L}(t)$ is the non-adiabatic part of interaction with an external field responsible for interband transitions,
\begin{equation*}
  \H{B} = \sum_{p,\vec q} \epsilon_{p\vec q} b_{p\vec q}^\dagger b_{p\vec q}
\end{equation*}
is the bath Hamiltonian, where %
the ladder operators $b_{p \vec q}^\dagger$ and $b_{p \vec q}$ satisfy the Bose commutation rules in the cases of phonons and many-particle states including an even number of carriers (excitons) and the Fermi commutation rules for odd number of carriers (defects with trapped charge carriers, trions).
The bath states are described by the quantum number $p$ and quasimomentum $\vec q$.
For simplicity of notation, we describe them by a single index $\alpha \equiv \{p, \vec q\}$.

In the case of the fast system and slow environment, the model Hamiltonian of system-bath interaction can be written as follows
\begin{equation}\label{e:HSB}
  \H{SB}(t) = \sum_{n, \vec k} \sum_{\alpha} S_{n\vec k,\alpha}(t) B_{\alpha}(t)
\end{equation}
where
\begin{subequations}\label{e:SB}
\begin{gather}
  S_{n\vec k,\alpha}(t) \equiv \frac{f_{n\vec k,\alpha}(t)}{\sqrt{2}} a_{n\vec k}^\dagger a_{n\vec k}^{\vphantom{\dagger}}
,\\
  B_{\alpha}(t) \equiv g_{\alpha} b_{\alpha}^\dagger + g_{\alpha}^* b_{\alpha}
\end{gather}
\end{subequations}
are the operators acting only on the system and bath states,
$g_{\alpha}$ is the amplitude of system-bath interaction determined by intrinsic bath properties, and
$f_{n\vec k,\alpha}(t)$ is the time-dependent part of the interaction amplitude depending on the rapidly changing parameters, e.g., charge carrier density or kinetic energy.

To obtain the equations of motion beyond the Markov approximation, we employ the time-convolutionless (TCL) projection operator technique~\cite{Tokuyama_1976_PTP_55_411, Ahn_1994_PRB_50_8310, Yamaguchi_2017_PRE_95_012136}, which yields the following equation for the reduced density matrix in the interaction representation
\begin{equation}\label{e:Redfield}
  \frac{d}{dt} \rh{S}(t) = -\int_0^t d t_1 \Tr_\mathrm{B}[\H{SB}(t), [\H{SB}(t_1), \rh{S}(t) \otimes \rh{B}]].
\end{equation}

Substituting~\eqref{e:HSB} into~\eqref{e:rhoS} and replacing the $t_1$ with $\tau = t-t_1$ in the integrand, one obtains
\begin{multline}\label{e:rhoS}
  \frac{d}{dt} \widetilde\rho_\mathrm{S}(t) = \sum_{n,m,\vec k} \sum_{\alpha,\beta} 
  \int_0^{\Delta t} d\tau\,
  \Bigl\{
    C_{\alpha\beta}  (\tau) [\widetilde S_{m\vec k,\beta}(t-\tau) \trh{S}(t), \widetilde S_{n\vec k,\alpha}^\dagger(t)]
\\
  + C_{\alpha\beta}^*(\tau) [\widetilde S_{n\vec k,\alpha}^\dagger(t), \trh{S}(t)\widetilde S_{m\vec k,\beta}(t-\tau)]
  \Bigr\},
\end{multline}
where $\Delta t = t - t_0$ is the time elapsed since the initial time $t_0$,
$C_{\alpha\beta}(\tau) \equiv \Tr_\mathrm{B} [\widetilde B_\alpha(\tau) \widetilde B_\beta(0)\rh{B}(0)]$ is the bath correlation function,
and $\widetilde{B}_\alpha(\tau) = g_\alpha b_\alpha^\dagger \ee^{i\w{\alpha} \tau}$.

In the Markov approximation, one assumes that the bath correlation function $C_{\alpha\beta}$ decays much faster than $\Delta t$, which allows extending the integration limit to infinity ($\Delta t \rightarrow \infty$).
We set $t_0 = 0$ and keep the finite limit of integration over $\tau$ to consider the non-Markovian case.
Eq.~\eqref{e:rhoS} is time-local only in terms of the density matrix $\rh{S}(t)$, but it is not in the Lindblad form because the convolution between $C_{\alpha\beta}(\tau)$ and $\widetilde S_{m\vec k,\beta}(t-\tau)$ is still required.
As shown below, in the case of pure dephasing, it can be reduced to the Lindblad form.

Following~\cite{Yamaguchi_2017_PRE_95_012136}, we rewrite $\widetilde S_{m\vec k}(t-\tau)$ as
\begin{equation*}
  \widetilde S_{m\vec k, \beta}(t-\tau) = \US^\dagger(t, t_0) \US(t,t-\tau) 
  S_{m\vec k, \beta}(t_0) \US^\dagger (t,t-\tau) \US(t,t_0),
\end{equation*}
where the evolution operator is separated in two parts.

The part describing the evolution from $t-\tau$ to $t$ can be approximated as
\begin{equation}\label{e:US}
  \US(t, t-\tau) \approx \exp\left[-i \H{S}(t) \tau\right].
\end{equation}

Assuming the completeness of the instantaneous eigenbasis, one can decompose the operator $\widetilde S_\beta(t)$ into summation over all instantaneous Bohr frequencies $\omega(t)$ in the system:
\begin{equation}
  \widetilde S_{m\vec k, \beta}(t) = \sum_{\omega(t)} \widetilde S_{m\vec k, \beta}(\omega(t)).
\end{equation}

This leads to the following approximation for the operator
\begin{equation}\label{e:tS}
  \widetilde S_{m\vec k, \beta}(t-\tau) \approx \sum_{\omega(t)} e^{i\omega(t) \tau} \widetilde S_{m\vec k, \beta}(\omega(t),t),
\end{equation}

\begin{multline}\label{e:RS}
  \mathcal R[\rh{S}(t)] = \sum_{n,m,\vec k} \sum_{\alpha,\beta} \sum_{\omega(t)} \gamma_{\alpha\beta}(\omega(t),t) 
  \Bigl\{
    S_{m\vec k, \beta}(\omega(t),t) \rh{S} S_{n\vec k, \alpha}^\dagger(\omega(t),t)
\\
  - S_{n\vec k, \beta}^\dagger(\omega(t),t) S_{m\vec k, \alpha}(\omega(t),t) \rh{S}\Bigr\}
+ \text{H.c.}
\end{multline}

Here, the convolution between $C_{\alpha\beta}(\tau)$ and $\widetilde S_{m\vec k, \beta}(t-\tau)$ is represented by the sum
\begin{equation*}
  \int_0^{t} d\tau\,C_{\alpha\beta}(\tau) \widetilde S_{m\vec k, \beta}(t - \tau) 
= \sum_{\omega(t)} \gamma_{\alpha\beta}(\omega(t), t) \widetilde S_{m\vec k}(\omega(t), t),
\end{equation*}
and
\begin{equation*}
  \gamma_{\alpha\beta}(\omega(t),t) = \int_0^t C_{\alpha\beta}(\tau)  e^{i \omega(t) \tau} d \tau
\end{equation*}
is the spectral correlation tensor connected with the correlation function via the finite Fourier transform.

Pure dephasing is dominated by elastic collisions described by energy-conserving terms with zero Bohr frequencies $\omega(t) = 0$.
Then from~\eqref{e:tS} it follows that $\widetilde S_{n\vec k, \alpha}(t - \tau) \approx \widetilde S_{n\vec k,\alpha}(0,t)$, the relaxation superoperator~\eqref{e:RS} takes the Lindblad form, and the spectral correlation tensor is connected with the correlation function simply via time integration
\begin{equation*}
  \gamma_{\alpha\beta}(0,t) = \int_0^{t} C_{\alpha\beta}(\tau)\,d \tau.
\end{equation*}

We neglect the non-diagonal elements of the correlation function and use a single index to enumerate the diagonal ones: $C_{\alpha} \equiv C_{\alpha\alpha}$.
Transforming Eq.~\eqref{e:rhoS} back from the interaction to the Schr\"odinger picture and using the definitions~\eqref{e:SB}, we obtain the master equation in the Redfield form:
\begin{equation}\label{e:rhonmk}
  \frac{d}{dt} \rho_{nm,\vec k}(t) = -i[H_\mathrm{S}, \rho]_{nm,\vec k} 
  - \gamma_{nm,\vec k}^{(\alpha)}(t) \rho_{nm,\vec k}(t),
\end{equation}
where 
\begin{equation}\label{e:gammanmk}
  \gamma_{nm,\vec k}^{(\alpha)}(t) = \frac{1}{2}[\gamma_{n\vec k, \alpha}(t) + \gamma_{m\vec k, \alpha}(t)]
,\quad
  \gamma_{n\vec k, \alpha}(t) = 4 |f_{n\vec k, \alpha}(t)|^2 \Re\int_0^{t} C_{\alpha}(\tau)\,d\tau,\quad
  n \ne m
\end{equation}
is the time-dependent pure dephasing rate.

Thus, the pure dephasing rate can be written as a product of the slowly varying envelope $G_\alpha(t) \equiv \Re \int_0^{t} C_\alpha(\tau)\,d\tau$ depending on the bath correlation function, and the rapidly varying function $F_{nm, \vec k}^{(\alpha)}(t) \equiv 2 \left\{|f_{n\vec k, \alpha}(t)|^2 + |f_{m\vec k, \alpha}(t)|^2\right\}$
\begin{equation}
  \gamma_{nm,\vec k}^{(\alpha)}(t) = G_\alpha(t) F_{nm, \vec k}^{(\alpha)}(t).
\end{equation}

Evolution of a quasiparticle interacting with an environment according to master equation~\eqref{e:rhonmk} can be described by the effective non-Hermitian Hamiltonian
\begin{equation*}
  H_\mathrm{eff}(t) = H_0 - \frac{i}{2}\gamma_{n\vec k}(t)
,
\end{equation*}
where the electron energies $E_{n\vec k}$ are replaced by the complex-valued time-dependent quasiparticle energies $\mathcal E_{n\vec k}(t)$.

Considering the field-matter interaction in the length gauge, one obtains the system of partial differential equations to the semiconductor Bloch equations~\cite{Schubert_2014_NP_8_119, Huttner_2017_LPR_11_1700049}.
Furthermore, applying the method of characteristics~\cite{Courant_1989} to the partial differential equations, one derives the Bloch acceleration theorem $\partial_t\vec K(t) = - \vec F(t)$ and the following system of ordinary differential equations similar to those in Refs.~\cite{Dunlap_1986_PRB_34_3625, Krieger_1986_PRB_33_5494, McDonald_2015_PRA_92_033845}:
\begin{subequations}\label{e:NMSBE}
\begin{align}
  \dot\rho_{nm, \vec K}(t) &= i\Omega_{nm, \vec K}^*(t) [\rho_{nn,\vec K}(t) - \rho_{mm,\vec K}(t)]
\notag\\
  &- i\sum_{l \ne n,m} [\Omega_{nl, \vec K}^*(t) \rho_{lm,\vec K} - \Omega_{lm}(t) \rho_{nl,\vec K}(t)]
,\\
  \dot\rho_{nn, \vec K}(t) &= i\sum_{l \ne n} \Omega_{nl, \vec K}(t) \rho_{nl,\vec K}(t) + \mathrm{c.c.}
\end{align}
\end{subequations}
where
\begin{equation}
  \Omega_{nm, \vec K}(t) = \vec F(t)\cdot \vec\xi_{nm,\vec K} \exp[i \phi_{nm,\vec K}(t)]
\end{equation}
is the matrix element of the field-matter interaction,
\begin{equation}
  \phi_{nm,\vec K}(t) = \int_{t_0}^{t} \Delta_{nm,\vec K}(t_1)\,d t_1,
\end{equation}
is the change of a total quantum phase between the Houston states in the bands $n$ and $m$,
$\Delta_{nm,\vec K}(t) = \mathcal E_{n\vec K}(t) - \mathcal E_{m\vec K}^*(t)$,
\begin{equation}
  \mathcal E_{n\vec K}(t) = E'_{n\vec K}(t) - \frac{i}{2} \gamma_{n\vec K}(t)
\end{equation}
is the quasiparticle energy describing the electron interacting with an environment,
$E'_{n\vec K}(t) = E_{n\vec K} + \vec F(t) \cdot \vec\xi_{nn,\vec K}$
are the modified band energies accounting for the geometric phase contribution.

Including the Coulomb interaction between electrons and keeping only the first-order terms, one obtains the well-known semiconductor Bloch equations~\cite{HaugKoch_2009} or the unscreened time-dependent Hartree--Fock approximation.
The semiconductor Bloch equations have the same form as~\eqref{e:NMSBE}, where the quasiparticle energies $\mathcal E_{n\vec K}$ and the interband matrix elements of the field-matter interaction $\Omega_{nm, \vec K}$ are renormalized by the Coulomb potential $V_{\vec q} = 1/q^2$~\cite{Lindberg_1988_PRB_38_3342, HaugKoch_2009, Huttner_2017_LPR_11_1700049}
\begin{gather}
\label{e:SigmaTDHF}
  \mathcal E_{n\vec K}^\mathrm{TDHF}(t) = E'_{n\vec K}(t)
  + \Sigma_{nn}^\mathrm{TDHF}(t) - i \frac{\gamma_{n\vec K}(t)}{2}
,\\
\label{e:OmegaHF}
  \Omega_{nm,\vec K}^\mathrm{TDHF}(t) = [ \vec F(t) \cdot \vec\xi_{nm,\vec K} - \Sigma_{nm}^\mathrm{TDHF}(t) ] \exp[i \phi_{nm,\vec K}^\mathrm{TDHF}(t)]
,\\
  \Sigma_{nm}^\mathrm{TDHF}(t) = -\sum_{\vec q\ne \vec K} V_{|\vec K - \vec q|} \rho_{nm,\vec q}(t)
\end{gather}

Note that the diagonal matrix elements of the self-energy $\Sigma_{nn}^\mathrm{TDHF}(t)$ are real-valued quantities.
Imaginary part of quasiparticle energy describing the damping of single-particle states due to interaction with the many-particle environment appears only in the higher-order approximations, e.g., the $GW$ or coupled-clusters, which are very computationally expensive and currently applicable for real-time simulations of simple atomic systems~\cite{Sato_2018_JCP_148_051101}.
Therefore, a reasonable non-Markovian model for $\gamma_{n\vec K}(t)$ is still required to describe pure dephasing.

\section{Bath of harmonic oscillators}

As was originally demonstrated by Feynman~\cite{Feynman_2000_AP_281_547}, Caldeira and Leggett~\cite{Caldeira_1983_AP_149_374}, the interaction with \textit{any} structured environment can be rigorously mapped onto a bath of harmonic oscillators, if the interaction is sufficiently weak and perturbation theory is applicable.
In this model, the bath is characterized by the spectral function
$$J(\omega) = \sum_\alpha |g_\alpha|^2 \delta(\omega - \omega_\alpha)$$
describing the distribution of oscillator's energy levels and their coupling to the system.

To simplify further analysis, we assume the ohmic spectral density with an exponential cutoff~\cite{Schlosshauer_2007, Breuer_2016_RMP_88_021002}
\begin{equation}
  J(\omega) = J_0 \omega e^{-\omega/\w{c}}
\end{equation}
allowing for analytical expressions of both the correlation function and relaxation rate.
Here, $J_0 = \gamma_0/\w{c}$ is a dimensionless constant,
$\gamma_0$ is the dephasing rate amplitude, and
$\w{c}$ is the cutoff energy.
For the many-particle environment, the cutoff is defined by the exciton binding energy $\Eex$, and for the phonon environment, it is given by the highest phonon energy $\ELO$.

If the bath in thermal equilibrium before interaction with the system and approaches it afterward, its correlation function is given by (Ref.~\cite{Schlosshauer_2007}, p. 181):
\begin{equation}\label{e:Cph}
  C(t, T) = \frac{1}{2\pi}\int_{-\infty}^{\infty} J(\omega) 
  \left[ \cos(\omega t) \coth\left(\frac{\omega}{2k_\mathrm{B} T}\right) - i\sin(\omega t) \right] d\omega,
\end{equation}
where the factor
\begin{equation*}
  \coth\left(\frac{\omega}{2k_\mathrm{B} T}\right) = 1 + 2N(\omega, T)
,\quad
  N(\omega, T) = \frac{1}{\ee^{\omega/\kB T} - 1}
\end{equation*}
appears due to the Bose--Einstein population distribution $N(\omega, T)$.
Here, the spectral function is extended to negative frequencies as $J(-\omega) = -J(\omega)$.

To make the integration in~\eqref{e:Cph} analytical, we assume that the thermal energy $\kB T_\alpha$ of the environment is much smaller than the cutoff frequency $\w{c}$.
For \aSiO{} at the room temperature ($T = 300$ K, $\kB T \approx 25$ meV), this condition is fully satisfied for both phonon ($E_\mathrm{LO} \approx 150$~meV) and many-particle environments ($E_\mathrm{ex} \approx 1.2$~eV).
Thus the dephasing rate envelope in the harmonic oscillator model is given by the following expression
\begin{gather}
\label{e:Galpha}
  G(t, T) \approx \gamma_0 \left\{
  \frac{2 \w{c} t}{1 + (\w{c} t)^2}
+ \frac{1}{\w{c} t} [\pi t k_\mathrm{B} T \coth(\pi t k_\mathrm{B} T) - 1]
  \right\}.
\end{gather}

In general, the dephasing rate envelope in the harmonic oscillator model is determined by two contributions: the first term of~\eqref{e:Galpha} describes quantum vacuum fluctuations, and the second term corresponds to thermal fluctuations.
Both of them strongly depend on the cutoff energy.

\begin{figure}
  \includegraphics[width=\linewidth]{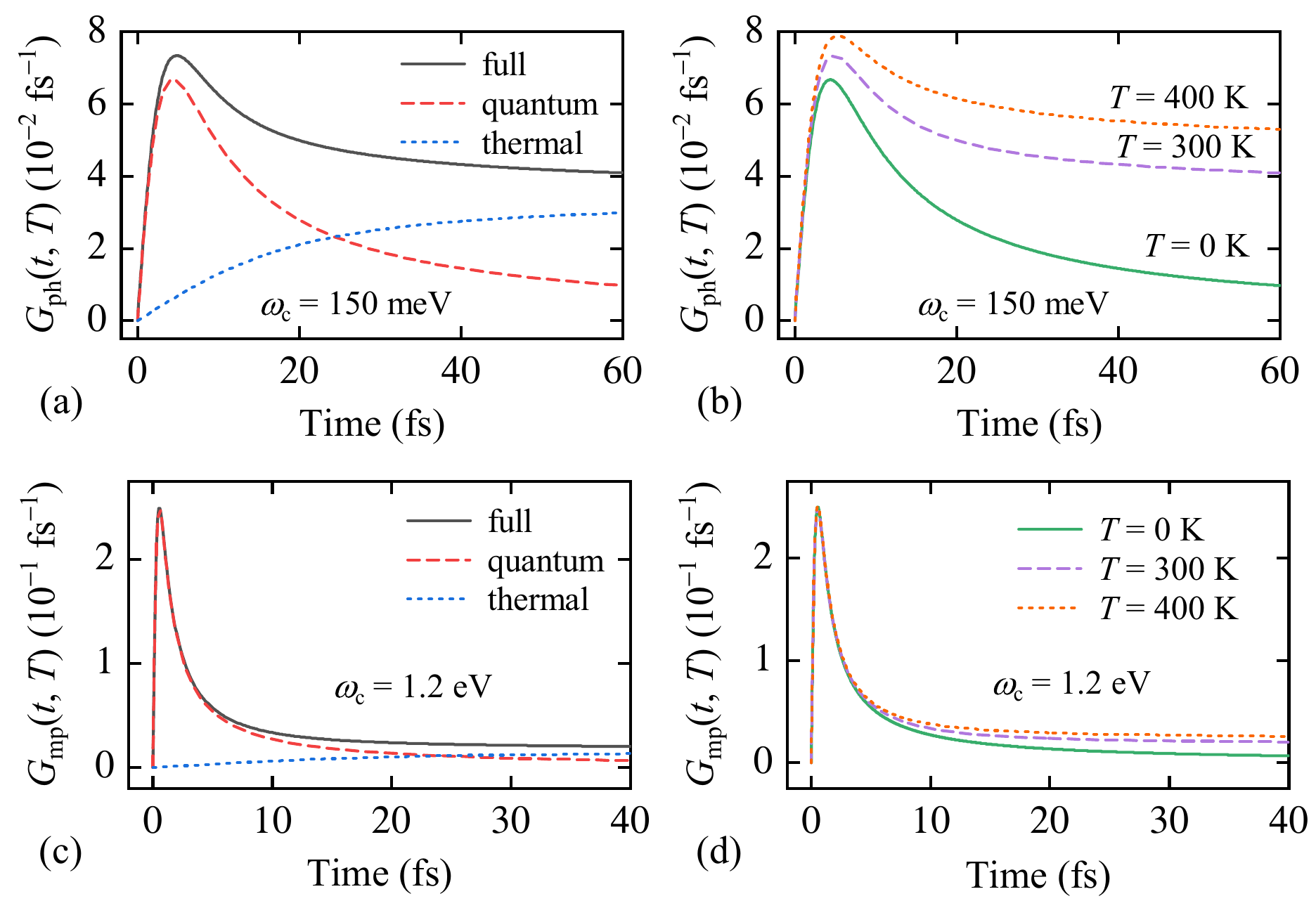}
  \caption{\label{f:ph_ex_rates}%
    (a) and (c) Time dependence of contributions from the quantum vacuum (dashed curve) and thermal fluctuations (dotted curve) to the full dephasing rate envelope $G_\alpha(t, T)$ (solid curve) at $T = 300$~K for the spectral function cutoffs $\w{c} = \ELO^{\aSiO} = 150$~meV and $\w{c} = \Eex^{\aSiO} = 1.2$~eV, respectively.
    (b)~and (d) Dephasing rates of the baths with phononic and excitonic cutoffs at three different temperatures.
  }
\end{figure}

Figures~\ref{f:ph_ex_rates}a and~\ref{f:ph_ex_rates}c show the dephasing rate envelopes calculated according to~\eqref{e:Galpha} for the cutoff given by the LO phonon and exciton binding energies, respectively.
In the short-time regime $t \leqslant 1/\w{c}$, the main contribution is given by the quantum vacuum fluctuations.
At times longer than the thermal correlation time $\ta{therm.} = 1/(\kB T) = 25.46$~fs, the contribution of thermal fluctuations term becomes dominant, and the quantum fluctuations vanish.
This effect is much more prominent for larger cutoff energies.

Figs.~\ref{f:ph_ex_rates}b and~\ref{f:ph_ex_rates}d compare time evolution of the rate envelopes for three different temperatures and two cutoffs.
In the long time limit ($t \rightarrow\infty$), the bath becomes completely thermalized and the dephasing rate envelope reaches its Markovian limit, $G^\mathrm{(M)}(T) = \gamma_0\pi \kB T/\w{c}$.
For the phonon bath, it is comparable to the peak value due to initial quantum fluctuations, but for the many-particle bath, it is smaller by more than an order of magnitude, as shown in Fig.~\ref{f:ph_ex_rates}d.
This property explains why the phonon bath is approximately Markovian even on a few-femtosecond time scale.
On the other hand, to reproduce the behavior of many-particle bath with the Markov approximation, one has to assume very short dephasing times $T_2$, which was done in previous simulations of high-harmonics spectroscopy~\cite{Luu_2015_Nature_521_498, Garg_2016_Nature_538_7625, Vampa_2014_PRL_113_073901}.

\begin{figure}
  \includegraphics[scale=0.95]{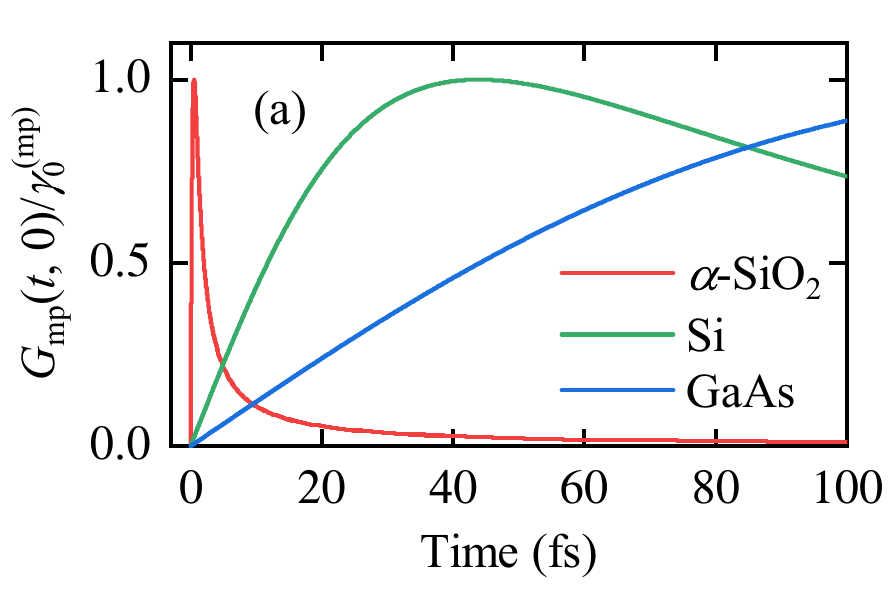}
  \includegraphics[scale=0.95]{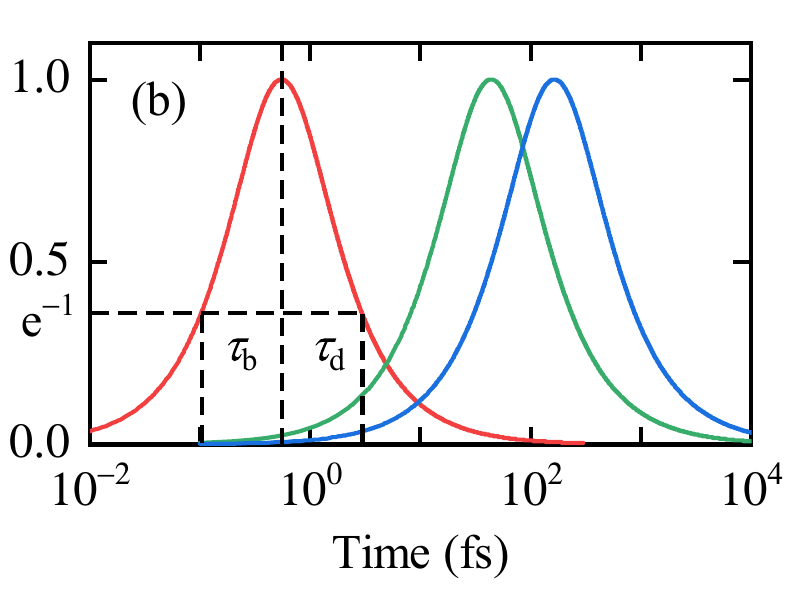}
  \caption{\label{f:correlation}%
    Comparison of the quantum vacuum fluctuation terms $G_\mathrm{mp}(t, 0)$ on the linear (a) and logarithmic (b) time axes.
    The spectral cutoff $\omega_\text{c}$ is given by the exciton binding energies of three representative materials: 
    $\Eex^{\aSiO} = 1.2$~eV,
    $\Eex^\text{Si} = 15$~meV, and
    $\Eex^\text{GaAs} = 4$~meV.
    Here, $\ta{b}$ and $\ta{d}$ are the build-up and decay times of the dephasing rate, respectively.
  }
\end{figure}

Figs.~\ref{f:correlation}a and~\ref{f:correlation}b compares the time dependencies of the quantum vacuum fluctuation terms, where the cutoff is determined by an exciton binding energies of a dielectric \aSiO{} and two semiconductors (Si and GaAs).
As shown in Fig.~\ref{f:correlation}b, the time-dependent dephasing rate can be characterized by the build-up $\tau_\mathrm{b}$ and decay $\tau_\mathrm{d}$ times at which it increases and decreases by $\ee$ times, respectively.
Very high exciton binding energy in a dielectric results in very fast dynamics with
$\tau_\mathrm{b}^{\aSiO} = 0.44$~fs and
$\tau_\mathrm{d}^{\aSiO} = 2.33$~fs.
For semiconductors, the bath evolves on a much slower time scales:
$\tau\mathrm{_b^{Si}} = 35.52$~fs,
$\tau\mathrm{_d^{Si}} = 186.32$~fs, and
$\tau\mathrm{_b^{GaAs}} = 133.19$~fs,
$\tau\mathrm{_d^{GaAs}} = 698.68$~fs.
To the best of our knowledge, these parameters were not measured experimentally, and thus, present an interest for further experimental investigations.

In the previous works on semiconductors exposed to terahertz laser pulses~\cite{Becker_1988_PRL_61_1647, Wang_1993_PRL_71_1261, Huegel_1999_PRL_83_3313}, the experimentally measured dephasing rate was well described within the excitation-induced dephasing (EID) model, where the scattering rate is inversely proportional to the mean inter-particle distance $\overline r(t) = \rho^{-1/3}(t)$
\begin{equation}\label{e:gammaEID}
  \gamma_{nm,\vec k}^\mathrm{EID}(t) = \gamma_0 + \gamma_1\rho^{1/3}(t).
\end{equation}
Here, $\rho(t)$ is the density of excited charge carriers, and $\gamma_0$ is the dephasing rate accounting for level broadening due to intrinsic lattice defects and frozen phonons, and $\gamma_1$ is the dephasing rate due to electron-electron scattering.

Equation~\eqref{e:gammaEID} provides a good fit of experimentally-observed dephasing rate in semiconductors excited by THz pulses and resembles other theoretical results, such as the Kohn-Sham-G\'asp\'ard exchange potential $V_\mathrm{KSG} \propto \rho^{1/3}(t)$~\cite{Bechstedt_2015}, p. 76, and the retarded self-energy of a charge carrier in a quasi-equilibrium electron-hole plasma $\Sigma^\mathrm{r}(t) \propto \rho^{1/3}(t)$~\cite{HaugKoch_2009}, p. 156.
The EID model can be obtained as a particular case of Eq.~\eqref{e:gammanmk}, where $F_{nm,\vec k}^\mathrm{(ph)}(t) = 1$, $F_{nm,\vec k}^\mathrm{(mp)}(t) = \rho^{1/3}(t)$ and $C_\alpha(t) = \gamma_\alpha \delta(t)$.

\begin{figure}
  \includegraphics[scale=1.0]{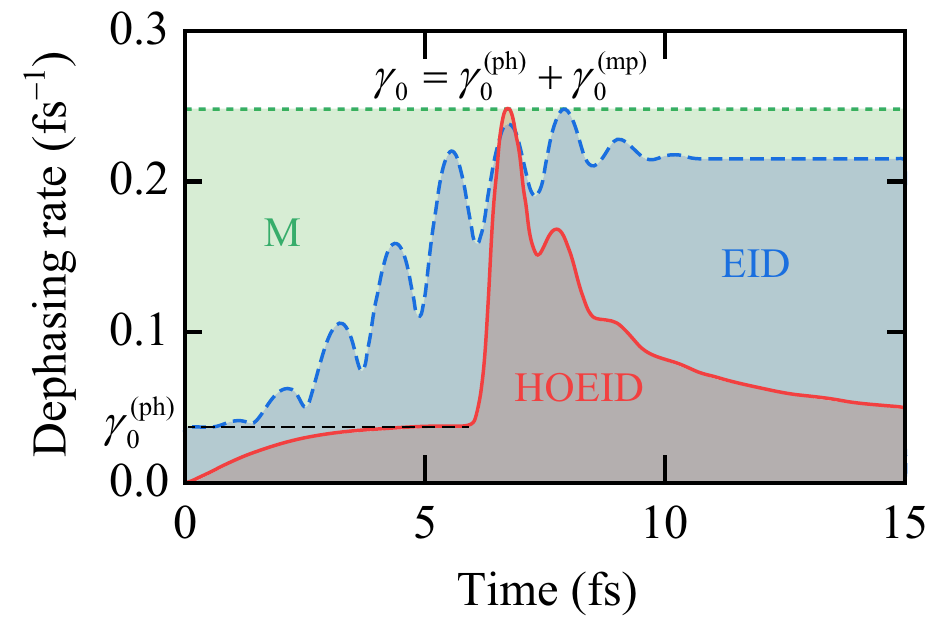}
  \caption{\label{f:deph_rates}%
    Dephasing rates of the Markovian (short-dashed line), EID (dashed curve) and HOEID (solid curve) models at the dephasing rate amplitude $\gamma_0 = 0.25$~fs$^{-1}$.
  }
\end{figure}

In Fig.~\ref{f:deph_rates}, we compare the Markovian and excitation-induced dephasing models with a constant (EID) and harmonic oscillator envelopes (HOEID).
The switch-on time for the many-particle environment is synchronized with the main optical cycle of the laser pulse, where the majority of charge carriers is excited, and response to the electric field becomes significantly nonlinear.
As one can see from Eqs.~\eqref{e:NMSBE}, the band populations are determined by the integral of dephasing rate.
The HOEID model has a smallest area under the curve, so it should have a lower excitation probability than the other models, while allowing for a temporally high dephasing rate.
This hypothesis will be numerically tested in the next section.

\section{Numerical results and discussion}

\begin{figure}[!htb]
  \includegraphics[width=\linewidth]{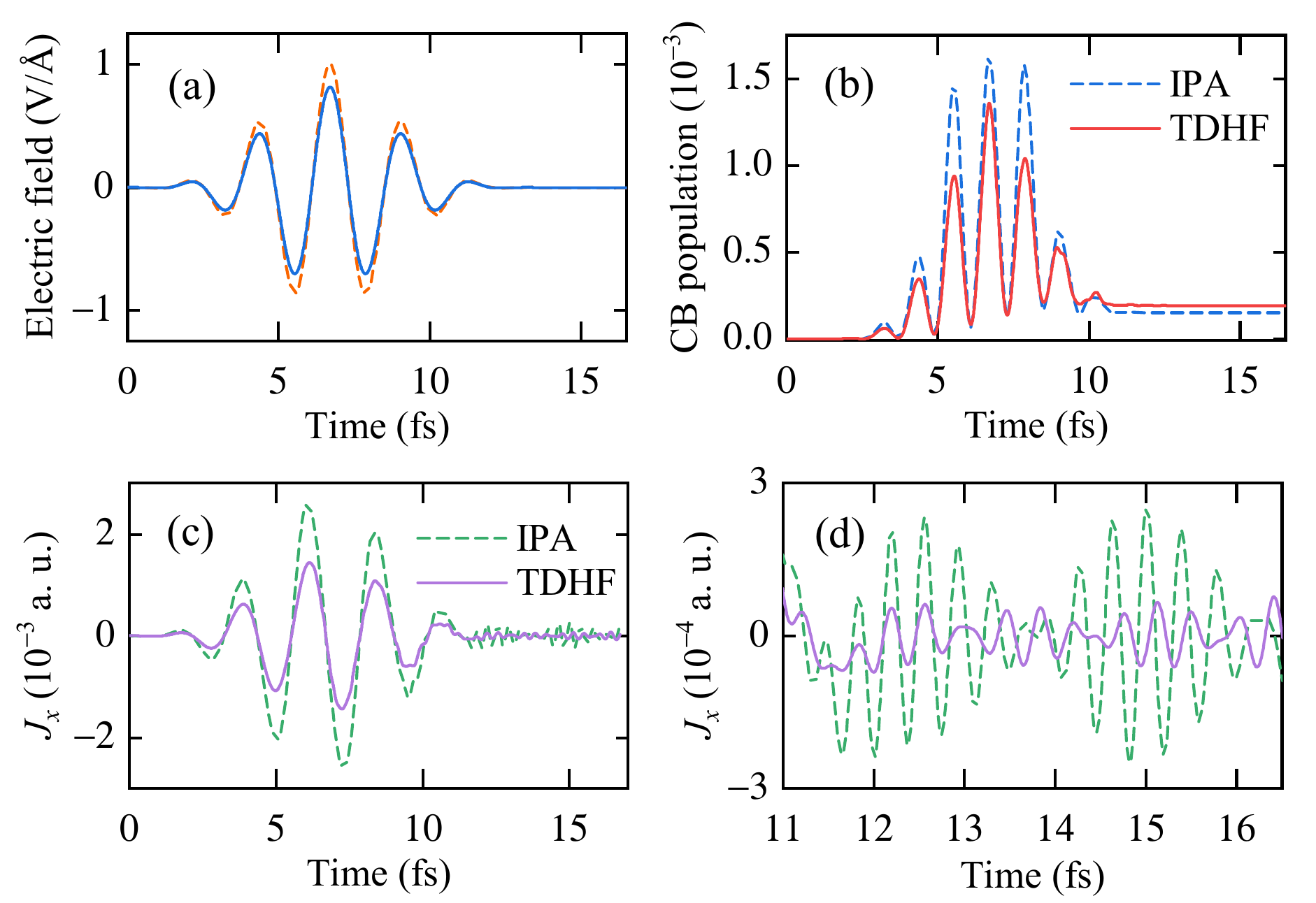}
  \caption{\label{f:IPA_TDHF}%
    Response of \aSiO{} to a strong near-infrared laser pulse without dephasing.
    (a)~Waveform of electric field with the $\cos^4$ envelope,
    $\omega_0 = 1.65$~eV, FWHM = 3.5 fs, $F_0 = 1$~V/\AA{}, in a vacuum (dashed line) and inside the medium (solid line).
    The screening factor is calculated from the Fresnel equation for a normal incidence~\cite{BornWolf_1980}.
    The field polarization is parallel to the $a$-axis of the crystal ($\Gamma-M$ direction in the reciprocal space)
    (b)~Time-dependent excitation probability calculated with the independent-particle (IPA) and the time-dependent Hartree--Fock (TDHF) approximations.
    (b)~Current density component $J_x$ along the $\Gamma-M$ direction in the Brillouin zone of the $\alpha$-quartz.
    (c)~Zoomed part of the current density after the laser pulse.
  }
\end{figure}

In this section, we present the numerical simulations for a bulk \aSiO{} interacting with the few-cycle IR laser pulse (Fig.~\ref{f:IPA_TDHF}a) to illustrate the main features of our non-Markovian dephasing model and compare it with other approximations.

Figures~\ref{f:IPA_TDHF}b--d show the comparison of simulations with equations in the independent-particle (IPA) and the time-dependent Hartree--Fock (TDHF) approximations.
Time propagation with the Crank-Nicolson scheme was performed on a grid of $25 \times 5 \times 5$ $\vec k$ points with four valence and four conduction bands.
Both models give qualitatively similar results, but the TDHF model reduces the interband current and transient populations due to the coupling of density matrix elements at different $\vec k$ points and renormalization of interband interaction energy~\eqref{e:OmegaHF}.
On the other hand, the residual population has increased by nearly 12\% in the TDHF simulation (see Fig.~\ref{f:IPA_TDHF}b).
After the laser pulse, the current density predicted by IPA simulation demonstrates the rephasing effect (see Figs.~\ref{f:IPA_TDHF}c and~\ref{f:IPA_TDHF}d).
The total interband coherence is partially restored and oscillates with a period of $\sim 2.5$~fs.
This effect is not observed in the TDHF simulation partially including electron-electron interaction.

\begin{figure}[!ht]
  \includegraphics[width=0.6\linewidth]{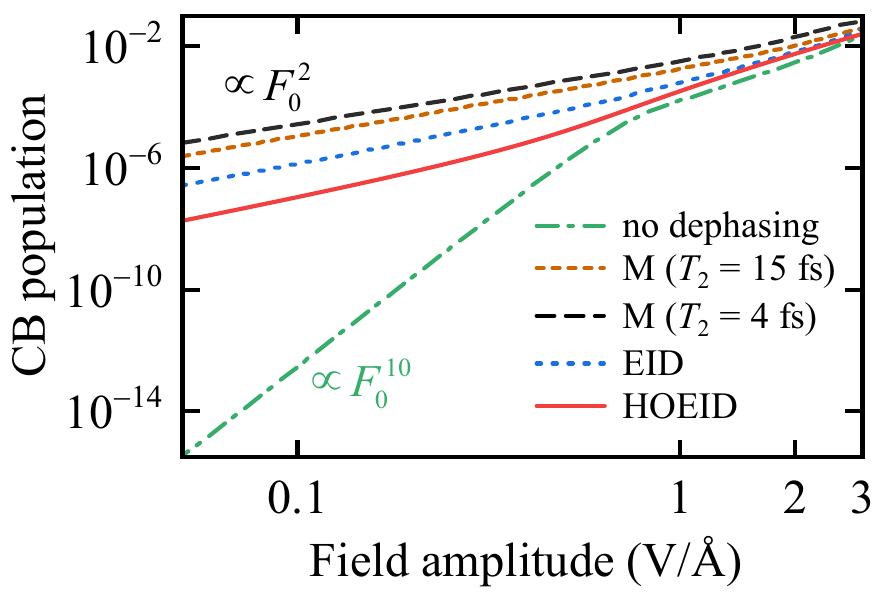}
  \caption{\label{f:amp_scaling}%
    Total conduction band population in $\alpha$-quartz after excitation by the IR laser pulse as a function of the field amplitude inside the solid for several representative models:
    fully coherent (dash-dotted curve),
    the Markov approximation with constant dephasing times $T_2 = 15$~fs (long-dashed line) and $T_2 = 4$~fs (short-dashed line),
    EID model with the constant envelope (dotted curve),
    and HOEID, which is the EID with time-dependent envelope given by the harmonic oscillator model) (solid curve).
    The pulse parameters are the same as in Fig.~\ref{f:IPA_TDHF}.
  }
\end{figure}

In Fig.~\ref{f:amp_scaling}, we compare numerical simulations of the charge carrier population on the amplitude of the NIR pulse with the $\cos^4$ envelope, where pure dephasing is described by the HOEID model (solid curve) with the other approximations: fully coherent TDHF equations, the EID model, and the Markovian constant decoherence rate.
As expected, the simulation without dephasing follows the perturbative scaling law $\propto F_0^{10}$ at low field amplitudes ($F_0 < 0.8$~V/\AA{}).
At higher fields, the scaling law changes due to closing of the lowest multiphoton channel predicted by the Keldysh theory and its modern generalization~\cite{Hawkins_2013_PRA_87_063842, Paasch-Colberg_2016_Optica_3_1358, Zhokhov_2014_PRL_113_133903, Shcheblanov_2017_PRA_96_063410}.
By contrast, the numerical simulation with a constant dephasing time $T_2 = 4$~fs, which is required for reproduction of the experimental HHG spectrum, shows a quadratic scaling of the population in entire range of the field amplitudes.
This corresponds to an artificial single-photon excitation channel due to spectral broadening, which is not typical for solids, but Ref.~\onlinecite{Schuh_2017_PRL_118_063901} demonstrates an opposite situation in gases, where only the Markovian dephasing introducing the single-photon excitation channel allows to reproduce the experimental observations.

As shown in Fig.~\ref{f:amp_scaling}, the unphysical scaling of excitation probability can be corrected by using the time-dependent dephasing rates.
The outcome of HOEID model approaches closer to the coherent one at high field amplitudes $F_0 \gtrsim 0.8$~V/\AA{}, where the ponderomotive energy becomes sufficiently large ($U_\mathrm{p} > \omega_0$) to overcome the spectral broadening introduced by pure dephasing.
In the recent CEP current control measurements~\cite{Chen_2018_NC_9_2070}, scaling of the transferred charge $Q \propto F_0^N$ with powers smaller than the perturbative result $N = 11$ were observed.
This observation can also be explained by a non-trivial dependence of pure dephasing rate on laser field and material parameters.
A rigorous analysis of similar measurements and the high-harmonic generation spectroscopy with simulations based on Eq.~\eqref{e:NMSBE} can be used for determining of the material-specific time-dependent dephasing rates.

\begin{figure}[!ht]
  \includegraphics[scale=1.0]{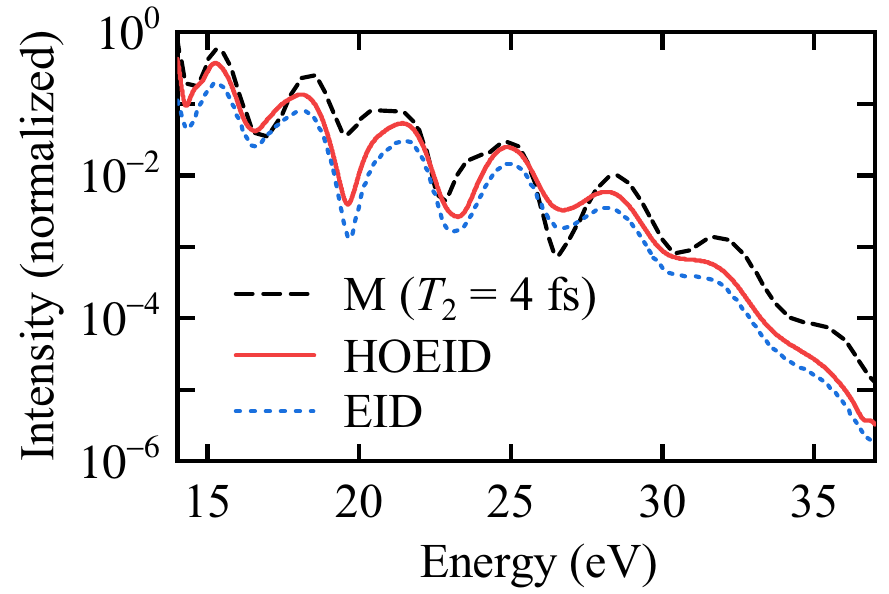}
  \caption{\label{f:HHG}%
    High harmonics spectra calculated with the constant and various time-dependent dephasing rates at $F_0 = 1$~V/\AA{}.}
\end{figure}

Finally, we compare the simulations of high harmonic spectra with different models of dephasing.
Fig.~\ref{f:HHG} shows that simulation obtained with the constant dephasing time $T_2 = 4$~fs (dashed curve) still gives the best agreement with experimental results, where the cutoff is extended beyond 35~eV~\cite{Luu_2015_Nature_521_498, Garg_2016_Nature_538_7625}.
The HOEID model approaches closer to the Markovian result than the EID.
Note that both the shape and intensity of high harmonics are sensitive to the time dependence of dephasing rate, which suggests that the high-harmonic spectroscopy can be used for reconstruction of time-dependent dephasing rates containing information on interaction with phonon and many-particle environments.

\section{Conclusions}

To summarize, we developed the non-Markovian theory of pure dephasing in a dielectric excited by an ultrashort IR/visible laser pulse.
It is shown that in the case of fast single-particle states and slow environment the adiabatic separation of system and bath results in the time-dependent dephasing rate with a slowly-varying envelope defined by the bath spectral function and rapidly-varying part determined by system's interaction with an external field.

We studied both phonon and many-particle baths within the harmonic oscillator model and showed that the spectral function cutoff significantly changes time-dependent envelope of the dephasing rate as well as its peak and thermalized values.
This explains why the phonon bath is approximately Markovian even on a few-femtosecond time scale and why the many-particle bath features unusually high values of the dephasing rate, which were not observed in the experiments with much longer laser pulses.

Numerical simulations show that the time-dependent dephasing rate with the envelope derived from the harmonic oscillator model significantly improves the problem of overestimated excitation probability at high intensities and allows for a temporally high dephasing rate, which is necessary for reproducing the experimental HHG spectrum of \aSiO{}.

\section{Acknowledgements}

I would like to acknowledge Prof. Georg Kresse for valuable discussions on usage and development of the VASP code.
This research was supported by the Austrian Science Fund (FWF) within the Lise Meitner Project No. M2198-N30.
The numerical calculations were partially performed at the Vienna Scientific Cluster (VSC-3).

\bibliographystyle{apsrev4-1}
\bibliography{article}

\begin{thebibliography}{65}%
\makeatletter
\providecommand \@ifxundefined [1]{%
 \@ifx{#1\undefined}
}%
\providecommand \@ifnum [1]{%
 \ifnum #1\expandafter \@firstoftwo
 \else \expandafter \@secondoftwo
 \fi
}%
\providecommand \@ifx [1]{%
 \ifx #1\expandafter \@firstoftwo
 \else \expandafter \@secondoftwo
 \fi
}%
\providecommand \natexlab [1]{#1}%
\providecommand \enquote  [1]{``#1''}%
\providecommand \bibnamefont  [1]{#1}%
\providecommand \bibfnamefont [1]{#1}%
\providecommand \citenamefont [1]{#1}%
\providecommand \href@noop [0]{\@secondoftwo}%
\providecommand \href [0]{\begingroup \@sanitize@url \@href}%
\providecommand \@href[1]{\@@startlink{#1}\@@href}%
\providecommand \@@href[1]{\endgroup#1\@@endlink}%
\providecommand \@sanitize@url [0]{\catcode `\\12\catcode `\$12\catcode
  `\&12\catcode `\#12\catcode `\^12\catcode `\_12\catcode `\%12\relax}%
\providecommand \@@startlink[1]{}%
\providecommand \@@endlink[0]{}%
\providecommand \url  [0]{\begingroup\@sanitize@url \@url }%
\providecommand \@url [1]{\endgroup\@href {#1}{\urlprefix }}%
\providecommand \urlprefix  [0]{URL }%
\providecommand \Eprint [0]{\href }%
\providecommand \doibase [0]{http://dx.doi.org/}%
\providecommand \selectlanguage [0]{\@gobble}%
\providecommand \bibinfo  [0]{\@secondoftwo}%
\providecommand \bibfield  [0]{\@secondoftwo}%
\providecommand \translation [1]{[#1]}%
\providecommand \BibitemOpen [0]{}%
\providecommand \bibitemStop [0]{}%
\providecommand \bibitemNoStop [0]{.\EOS\space}%
\providecommand \EOS [0]{\spacefactor3000\relax}%
\providecommand \BibitemShut  [1]{\csname bibitem#1\endcsname}%
\let\auto@bib@innerbib\@empty
\bibitem [{\citenamefont {Goulielmakis}\ \emph {et~al.}(2008)\citenamefont
  {Goulielmakis}, \citenamefont {Schultze}, \citenamefont {Hofstetter},
  \citenamefont {Yakovlev}, \citenamefont {Gagnon}, \citenamefont {Uiberacker},
  \citenamefont {Aquila}, \citenamefont {Gullikson}, \citenamefont {Attwood},
  \citenamefont {Kienberger}, \citenamefont {Krausz},\ and\ \citenamefont
  {Kleineberg}}]{Goulielmakis_2008_SCI_320_1614}%
  \BibitemOpen
  \bibfield  {author} {\bibinfo {author} {\bibfnamefont {E.}~\bibnamefont
  {Goulielmakis}}, \bibinfo {author} {\bibfnamefont {M.}~\bibnamefont
  {Schultze}}, \bibinfo {author} {\bibfnamefont {M.}~\bibnamefont
  {Hofstetter}}, \bibinfo {author} {\bibfnamefont {V.~S.}\ \bibnamefont
  {Yakovlev}}, \bibinfo {author} {\bibfnamefont {J.}~\bibnamefont {Gagnon}},
  \bibinfo {author} {\bibfnamefont {M.}~\bibnamefont {Uiberacker}}, \bibinfo
  {author} {\bibfnamefont {A.~L.}\ \bibnamefont {Aquila}}, \bibinfo {author}
  {\bibfnamefont {E.~M.}\ \bibnamefont {Gullikson}}, \bibinfo {author}
  {\bibfnamefont {D.~T.}\ \bibnamefont {Attwood}}, \bibinfo {author}
  {\bibfnamefont {R.}~\bibnamefont {Kienberger}}, \bibinfo {author}
  {\bibfnamefont {F.}~\bibnamefont {Krausz}}, \ and\ \bibinfo {author}
  {\bibfnamefont {U.}~\bibnamefont {Kleineberg}},\ }\href {\doibase
  10.1126/science.1157846} {\bibfield  {journal} {\bibinfo  {journal}
  {Science}\ }\textbf {\bibinfo {volume} {320}},\ \bibinfo {pages} {1614}
  (\bibinfo {year} {2008})}\BibitemShut {NoStop}%
\bibitem [{\citenamefont {Huang}\ \emph {et~al.}(2011)\citenamefont {Huang},
  \citenamefont {Cirmi}, \citenamefont {Moses}, \citenamefont {Hong},
  \citenamefont {Bhardwaj}, \citenamefont {Birge}, \citenamefont {Chen},
  \citenamefont {Li}, \citenamefont {Eggleton}, \citenamefont {Cerullo},\ and\
  \citenamefont {Kartner}}]{Huang_2011_NP_5_475}%
  \BibitemOpen
  \bibfield  {author} {\bibinfo {author} {\bibfnamefont {S.-W.}\ \bibnamefont
  {Huang}}, \bibinfo {author} {\bibfnamefont {G.}~\bibnamefont {Cirmi}},
  \bibinfo {author} {\bibfnamefont {J.}~\bibnamefont {Moses}}, \bibinfo
  {author} {\bibfnamefont {K.-H.}\ \bibnamefont {Hong}}, \bibinfo {author}
  {\bibfnamefont {S.}~\bibnamefont {Bhardwaj}}, \bibinfo {author}
  {\bibfnamefont {J.~R.}\ \bibnamefont {Birge}}, \bibinfo {author}
  {\bibfnamefont {L.-J.}\ \bibnamefont {Chen}}, \bibinfo {author}
  {\bibfnamefont {E.}~\bibnamefont {Li}}, \bibinfo {author} {\bibfnamefont
  {B.~J.}\ \bibnamefont {Eggleton}}, \bibinfo {author} {\bibfnamefont
  {G.}~\bibnamefont {Cerullo}}, \ and\ \bibinfo {author} {\bibfnamefont
  {F.~X.}\ \bibnamefont {Kartner}},\ }\href {\doibase 10.1038/nphoton.2011.140}
  {\bibfield  {journal} {\bibinfo  {journal} {Nat. Photon.}\ }\textbf {\bibinfo
  {volume} {5}},\ \bibinfo {pages} {475} (\bibinfo {year} {2011})}\BibitemShut
  {NoStop}%
\bibitem [{\citenamefont {Fattahi}\ \emph {et~al.}(2014)\citenamefont
  {Fattahi}, \citenamefont {Barros}, \citenamefont {Gorjan}, \citenamefont
  {Nubbemeyer}, \citenamefont {Alsaif}, \citenamefont {Teisset}, \citenamefont
  {Schultze}, \citenamefont {Prinz}, \citenamefont {Haefner}, \citenamefont
  {Ueffing}, \citenamefont {Alismail}, \citenamefont {V\'amos}, \citenamefont
  {Schwarz}, \citenamefont {Pronin}, \citenamefont {Brons}, \citenamefont
  {Geng}, \citenamefont {Arisholm}, \citenamefont {Ciappina}, \citenamefont
  {Yakovlev}, \citenamefont {Kim}, \citenamefont {Azzeer}, \citenamefont
  {Karpowicz}, \citenamefont {Sutter}, \citenamefont {Major}, \citenamefont
  {Metzger},\ and\ \citenamefont {Krausz}}]{Fattahi_2014_Optica_1_45}%
  \BibitemOpen
  \bibfield  {author} {\bibinfo {author} {\bibfnamefont {H.}~\bibnamefont
  {Fattahi}}, \bibinfo {author} {\bibfnamefont {H.~G.}\ \bibnamefont {Barros}},
  \bibinfo {author} {\bibfnamefont {M.}~\bibnamefont {Gorjan}}, \bibinfo
  {author} {\bibfnamefont {T.}~\bibnamefont {Nubbemeyer}}, \bibinfo {author}
  {\bibfnamefont {B.}~\bibnamefont {Alsaif}}, \bibinfo {author} {\bibfnamefont
  {C.~Y.}\ \bibnamefont {Teisset}}, \bibinfo {author} {\bibfnamefont
  {M.}~\bibnamefont {Schultze}}, \bibinfo {author} {\bibfnamefont
  {S.}~\bibnamefont {Prinz}}, \bibinfo {author} {\bibfnamefont
  {M.}~\bibnamefont {Haefner}}, \bibinfo {author} {\bibfnamefont
  {M.}~\bibnamefont {Ueffing}}, \bibinfo {author} {\bibfnamefont
  {A.}~\bibnamefont {Alismail}}, \bibinfo {author} {\bibfnamefont
  {L.}~\bibnamefont {V\'amos}}, \bibinfo {author} {\bibfnamefont
  {A.}~\bibnamefont {Schwarz}}, \bibinfo {author} {\bibfnamefont
  {O.}~\bibnamefont {Pronin}}, \bibinfo {author} {\bibfnamefont
  {J.}~\bibnamefont {Brons}}, \bibinfo {author} {\bibfnamefont {X.~T.}\
  \bibnamefont {Geng}}, \bibinfo {author} {\bibfnamefont {G.}~\bibnamefont
  {Arisholm}}, \bibinfo {author} {\bibfnamefont {M.}~\bibnamefont {Ciappina}},
  \bibinfo {author} {\bibfnamefont {V.~S.}\ \bibnamefont {Yakovlev}}, \bibinfo
  {author} {\bibfnamefont {D.-E.}\ \bibnamefont {Kim}}, \bibinfo {author}
  {\bibfnamefont {A.~M.}\ \bibnamefont {Azzeer}}, \bibinfo {author}
  {\bibfnamefont {N.}~\bibnamefont {Karpowicz}}, \bibinfo {author}
  {\bibfnamefont {D.}~\bibnamefont {Sutter}}, \bibinfo {author} {\bibfnamefont
  {Z.}~\bibnamefont {Major}}, \bibinfo {author} {\bibfnamefont
  {T.}~\bibnamefont {Metzger}}, \ and\ \bibinfo {author} {\bibfnamefont
  {F.}~\bibnamefont {Krausz}},\ }\href {\doibase 10.1364/OPTICA.1.000045}
  {\bibfield  {journal} {\bibinfo  {journal} {Optica}\ }\textbf {\bibinfo
  {volume} {1}},\ \bibinfo {pages} {45} (\bibinfo {year} {2014})}\BibitemShut
  {NoStop}%
\bibitem [{\citenamefont {Schiffrin}\ \emph {et~al.}(2013)\citenamefont
  {Schiffrin}, \citenamefont {Paasch-Colberg}, \citenamefont {Karpowicz},
  \citenamefont {Apalkov}, \citenamefont {Gerster}, \citenamefont
  {M\"uhlbrandt}, \citenamefont {Korbman}, \citenamefont {Reichert},
  \citenamefont {Schultze}, \citenamefont {Holzner}, \citenamefont {Barth},
  \citenamefont {Kienberger}, \citenamefont {Ernstorfer}, \citenamefont
  {Yakovlev}, \citenamefont {Stockman},\ and\ \citenamefont
  {Krausz}}]{Schiffrin_2013_Nature_493_70}%
  \BibitemOpen
  \bibfield  {author} {\bibinfo {author} {\bibfnamefont {A.}~\bibnamefont
  {Schiffrin}}, \bibinfo {author} {\bibfnamefont {T.}~\bibnamefont
  {Paasch-Colberg}}, \bibinfo {author} {\bibfnamefont {N.}~\bibnamefont
  {Karpowicz}}, \bibinfo {author} {\bibfnamefont {V.}~\bibnamefont {Apalkov}},
  \bibinfo {author} {\bibfnamefont {D.}~\bibnamefont {Gerster}}, \bibinfo
  {author} {\bibfnamefont {S.}~\bibnamefont {M\"uhlbrandt}}, \bibinfo {author}
  {\bibfnamefont {M.}~\bibnamefont {Korbman}}, \bibinfo {author} {\bibfnamefont
  {J.}~\bibnamefont {Reichert}}, \bibinfo {author} {\bibfnamefont
  {M.}~\bibnamefont {Schultze}}, \bibinfo {author} {\bibfnamefont
  {S.}~\bibnamefont {Holzner}}, \bibinfo {author} {\bibfnamefont {J.~V.}\
  \bibnamefont {Barth}}, \bibinfo {author} {\bibfnamefont {R.}~\bibnamefont
  {Kienberger}}, \bibinfo {author} {\bibfnamefont {R.}~\bibnamefont
  {Ernstorfer}}, \bibinfo {author} {\bibfnamefont {V.~S.}\ \bibnamefont
  {Yakovlev}}, \bibinfo {author} {\bibfnamefont {M.~I.}\ \bibnamefont
  {Stockman}}, \ and\ \bibinfo {author} {\bibfnamefont {F.}~\bibnamefont
  {Krausz}},\ }\href {\doibase 10.1038/nature11567} {\bibfield  {journal}
  {\bibinfo  {journal} {Nature}\ }\textbf {\bibinfo {volume} {493}},\ \bibinfo
  {pages} {70} (\bibinfo {year} {2013})}\BibitemShut {NoStop}%
\bibitem [{\citenamefont {Luu}\ \emph {et~al.}(2015)\citenamefont {Luu},
  \citenamefont {Garg}, \citenamefont {Kruchinin}, \citenamefont {Moulet},
  \citenamefont {Hassan},\ and\ \citenamefont
  {Goulielmakis}}]{Luu_2015_Nature_521_498}%
  \BibitemOpen
  \bibfield  {author} {\bibinfo {author} {\bibfnamefont {T.~T.}\ \bibnamefont
  {Luu}}, \bibinfo {author} {\bibfnamefont {M.}~\bibnamefont {Garg}}, \bibinfo
  {author} {\bibfnamefont {S.~Y.}\ \bibnamefont {Kruchinin}}, \bibinfo {author}
  {\bibfnamefont {A.}~\bibnamefont {Moulet}}, \bibinfo {author} {\bibfnamefont
  {M.~T.}\ \bibnamefont {Hassan}}, \ and\ \bibinfo {author} {\bibfnamefont
  {E.}~\bibnamefont {Goulielmakis}},\ }\href {\doibase 10.1038/nature14456}
  {\bibfield  {journal} {\bibinfo  {journal} {Nature}\ }\textbf {\bibinfo
  {volume} {521}},\ \bibinfo {pages} {498} (\bibinfo {year}
  {2015})}\BibitemShut {NoStop}%
\bibitem [{\citenamefont {Liu}\ \emph {et~al.}(2017)\citenamefont {Liu},
  \citenamefont {Li}, \citenamefont {You}, \citenamefont {Ghimire},
  \citenamefont {Heinz},\ and\ \citenamefont {Reis}}]{Liu_2017_NP_13_262}%
  \BibitemOpen
  \bibfield  {author} {\bibinfo {author} {\bibfnamefont {H.}~\bibnamefont
  {Liu}}, \bibinfo {author} {\bibfnamefont {Y.}~\bibnamefont {Li}}, \bibinfo
  {author} {\bibfnamefont {Y.~S.}\ \bibnamefont {You}}, \bibinfo {author}
  {\bibfnamefont {S.}~\bibnamefont {Ghimire}}, \bibinfo {author} {\bibfnamefont
  {T.~F.}\ \bibnamefont {Heinz}}, \ and\ \bibinfo {author} {\bibfnamefont
  {D.~A.}\ \bibnamefont {Reis}},\ }\href {\doibase 10.1038/nphys3946}
  {\bibfield  {journal} {\bibinfo  {journal} {Nat. Phys.}\ }\textbf {\bibinfo
  {volume} {13}},\ \bibinfo {pages} {262} (\bibinfo {year} {2017})}\BibitemShut
  {NoStop}%
\bibitem [{\citenamefont {Kelardeh}\ \emph {et~al.}(2016)\citenamefont
  {Kelardeh}, \citenamefont {Apalkov},\ and\ \citenamefont
  {Stockman}}]{Kelardeh_2016_PRB_93_155434}%
  \BibitemOpen
  \bibfield  {author} {\bibinfo {author} {\bibfnamefont {H.~K.}\ \bibnamefont
  {Kelardeh}}, \bibinfo {author} {\bibfnamefont {V.}~\bibnamefont {Apalkov}}, \
  and\ \bibinfo {author} {\bibfnamefont {M.~I.}\ \bibnamefont {Stockman}},\
  }\href {\doibase 10.1103/PhysRevB.93.155434} {\bibfield  {journal} {\bibinfo
  {journal} {Phys. Rev. B}\ }\textbf {\bibinfo {volume} {93}},\ \bibinfo
  {pages} {155434} (\bibinfo {year} {2016})}\BibitemShut {NoStop}%
\bibitem [{\citenamefont {Oliaei~Motlagh}\ \emph {et~al.}(2018)\citenamefont
  {Oliaei~Motlagh}, \citenamefont {Wu}, \citenamefont {Apalkov},\ and\
  \citenamefont {Stockman}}]{Motlagh_2018_PRB_98_125410}%
  \BibitemOpen
  \bibfield  {author} {\bibinfo {author} {\bibfnamefont {S.~A.}\ \bibnamefont
  {Oliaei~Motlagh}}, \bibinfo {author} {\bibfnamefont {J.-S.}\ \bibnamefont
  {Wu}}, \bibinfo {author} {\bibfnamefont {V.}~\bibnamefont {Apalkov}}, \ and\
  \bibinfo {author} {\bibfnamefont {M.~I.}\ \bibnamefont {Stockman}},\ }\href
  {\doibase 10.1103/PhysRevB.98.125410} {\bibfield  {journal} {\bibinfo
  {journal} {Phys. Rev. B}\ }\textbf {\bibinfo {volume} {98}},\ \bibinfo
  {pages} {125410} (\bibinfo {year} {2018})}\BibitemShut {NoStop}%
\bibitem [{\citenamefont {Han}\ \emph {et~al.}(2016)\citenamefont {Han},
  \citenamefont {Kim}, \citenamefont {Kim}, \citenamefont {Kim}, \citenamefont
  {Kim}, \citenamefont {Park},\ and\ \citenamefont
  {Kim}}]{Han_2016_NC_7_13105}%
  \BibitemOpen
  \bibfield  {author} {\bibinfo {author} {\bibfnamefont {S.}~\bibnamefont
  {Han}}, \bibinfo {author} {\bibfnamefont {H.}~\bibnamefont {Kim}}, \bibinfo
  {author} {\bibfnamefont {Y.~W.}\ \bibnamefont {Kim}}, \bibinfo {author}
  {\bibfnamefont {Y.-J.}\ \bibnamefont {Kim}}, \bibinfo {author} {\bibfnamefont
  {S.}~\bibnamefont {Kim}}, \bibinfo {author} {\bibfnamefont {I.-Y.}\
  \bibnamefont {Park}}, \ and\ \bibinfo {author} {\bibfnamefont {S.-W.}\
  \bibnamefont {Kim}},\ }\href {\doibase 10.1038/ncomms13105} {\bibfield
  {journal} {\bibinfo  {journal} {Nat. Comm.}\ }\textbf {\bibinfo {volume}
  {7}},\ \bibinfo {pages} {13105} (\bibinfo {year} {2016})}\BibitemShut
  {NoStop}%
\bibitem [{\citenamefont {Vampa}\ \emph {et~al.}(2017)\citenamefont {Vampa},
  \citenamefont {Ghamsari}, \citenamefont {Siadat~Mousavi}, \citenamefont
  {Hammond}, \citenamefont {Olivieri}, \citenamefont {Lisicka-Skrek},
  \citenamefont {Naumov}, \citenamefont {Villeneuve}, \citenamefont {Staudte},
  \citenamefont {Berini},\ and\ \citenamefont {Corkum}}]{Vampa_2017_NP_13_659}%
  \BibitemOpen
  \bibfield  {author} {\bibinfo {author} {\bibfnamefont {G.}~\bibnamefont
  {Vampa}}, \bibinfo {author} {\bibfnamefont {B.~G.}\ \bibnamefont {Ghamsari}},
  \bibinfo {author} {\bibfnamefont {S.}~\bibnamefont {Siadat~Mousavi}},
  \bibinfo {author} {\bibfnamefont {T.~J.}\ \bibnamefont {Hammond}}, \bibinfo
  {author} {\bibfnamefont {A.}~\bibnamefont {Olivieri}}, \bibinfo {author}
  {\bibfnamefont {E.}~\bibnamefont {Lisicka-Skrek}}, \bibinfo {author}
  {\bibfnamefont {A.~Y.}\ \bibnamefont {Naumov}}, \bibinfo {author}
  {\bibfnamefont {D.~M.}\ \bibnamefont {Villeneuve}}, \bibinfo {author}
  {\bibfnamefont {A.}~\bibnamefont {Staudte}}, \bibinfo {author} {\bibfnamefont
  {P.}~\bibnamefont {Berini}}, \ and\ \bibinfo {author} {\bibfnamefont {P.~B.}\
  \bibnamefont {Corkum}},\ }\href {\doibase 10.1038/nphys4087} {\bibfield
  {journal} {\bibinfo  {journal} {Nat. Phys.}\ }\textbf {\bibinfo {volume}
  {13}},\ \bibinfo {pages} {659} (\bibinfo {year} {2017})}\BibitemShut
  {NoStop}%
\bibitem [{\citenamefont {Ghimire}\ \emph {et~al.}(2011)\citenamefont
  {Ghimire}, \citenamefont {DiChiara}, \citenamefont {Sistrunk}, \citenamefont
  {Agostini}, \citenamefont {DiMauro},\ and\ \citenamefont
  {Reis}}]{Ghimire_2011_NP_7_138}%
  \BibitemOpen
  \bibfield  {author} {\bibinfo {author} {\bibfnamefont {S.}~\bibnamefont
  {Ghimire}}, \bibinfo {author} {\bibfnamefont {A.~D.}\ \bibnamefont
  {DiChiara}}, \bibinfo {author} {\bibfnamefont {E.}~\bibnamefont {Sistrunk}},
  \bibinfo {author} {\bibfnamefont {P.}~\bibnamefont {Agostini}}, \bibinfo
  {author} {\bibfnamefont {L.~F.}\ \bibnamefont {DiMauro}}, \ and\ \bibinfo
  {author} {\bibfnamefont {D.~A.}\ \bibnamefont {Reis}},\ }\href {\doibase
  10.1038/nphys1847} {\bibfield  {journal} {\bibinfo  {journal} {Nature
  Physics}\ }\textbf {\bibinfo {volume} {7}},\ \bibinfo {pages} {138} (\bibinfo
  {year} {2011})}\BibitemShut {NoStop}%
\bibitem [{\citenamefont {Vampa}\ \emph {et~al.}(2014)\citenamefont {Vampa},
  \citenamefont {McDonald}, \citenamefont {Orlando}, \citenamefont {Klug},
  \citenamefont {Corkum},\ and\ \citenamefont
  {Brabec}}]{Vampa_2014_PRL_113_073901}%
  \BibitemOpen
  \bibfield  {author} {\bibinfo {author} {\bibfnamefont {G.}~\bibnamefont
  {Vampa}}, \bibinfo {author} {\bibfnamefont {C.~R.}\ \bibnamefont {McDonald}},
  \bibinfo {author} {\bibfnamefont {G.}~\bibnamefont {Orlando}}, \bibinfo
  {author} {\bibfnamefont {D.}~\bibnamefont {Klug}}, \bibinfo {author}
  {\bibfnamefont {P.}~\bibnamefont {Corkum}}, \ and\ \bibinfo {author}
  {\bibfnamefont {T.}~\bibnamefont {Brabec}},\ }\href {\doibase
  10.1103/PhysRevLett.113.073901} {\bibfield  {journal} {\bibinfo  {journal}
  {Phys. Rev. Lett.}\ }\textbf {\bibinfo {volume} {113}},\ \bibinfo {pages}
  {073901} (\bibinfo {year} {2014})}\BibitemShut {NoStop}%
\bibitem [{\citenamefont {Garg}\ \emph {et~al.}(2016)\citenamefont {Garg},
  \citenamefont {Zhan}, \citenamefont {Luu}, \citenamefont {Lakhotia},
  \citenamefont {Klostermann}, \citenamefont {Guggenmos},\ and\ \citenamefont
  {Goulielmakis}}]{Garg_2016_Nature_538_7625}%
  \BibitemOpen
  \bibfield  {author} {\bibinfo {author} {\bibfnamefont {M.}~\bibnamefont
  {Garg}}, \bibinfo {author} {\bibfnamefont {M.}~\bibnamefont {Zhan}}, \bibinfo
  {author} {\bibfnamefont {T.~T.}\ \bibnamefont {Luu}}, \bibinfo {author}
  {\bibfnamefont {H.}~\bibnamefont {Lakhotia}}, \bibinfo {author}
  {\bibfnamefont {T.}~\bibnamefont {Klostermann}}, \bibinfo {author}
  {\bibfnamefont {A.}~\bibnamefont {Guggenmos}}, \ and\ \bibinfo {author}
  {\bibfnamefont {E.}~\bibnamefont {Goulielmakis}},\ }\href {\doibase
  10.1038/nature19821} {\bibfield  {journal} {\bibinfo  {journal} {Nature}\
  }\textbf {\bibinfo {volume} {538}},\ \bibinfo {pages} {359} (\bibinfo {year}
  {2016})}\BibitemShut {NoStop}%
\bibitem [{\citenamefont {Ndabashimiye}\ \emph {et~al.}(2016)\citenamefont
  {Ndabashimiye}, \citenamefont {Ghimire}, \citenamefont {Wu}, \citenamefont
  {Browne}, \citenamefont {Schafer}, \citenamefont {Gaarde},\ and\
  \citenamefont {Reis}}]{Ndabashimiye_2016_Nature_534_520}%
  \BibitemOpen
  \bibfield  {author} {\bibinfo {author} {\bibfnamefont {G.}~\bibnamefont
  {Ndabashimiye}}, \bibinfo {author} {\bibfnamefont {S.}~\bibnamefont
  {Ghimire}}, \bibinfo {author} {\bibfnamefont {M.}~\bibnamefont {Wu}},
  \bibinfo {author} {\bibfnamefont {D.~A.}\ \bibnamefont {Browne}}, \bibinfo
  {author} {\bibfnamefont {K.~J.}\ \bibnamefont {Schafer}}, \bibinfo {author}
  {\bibfnamefont {M.~B.}\ \bibnamefont {Gaarde}}, \ and\ \bibinfo {author}
  {\bibfnamefont {D.~A.}\ \bibnamefont {Reis}},\ }\href {\doibase
  10.1038/nature17660} {\bibfield  {journal} {\bibinfo  {journal} {Nature}\
  }\textbf {\bibinfo {volume} {534}},\ \bibinfo {pages} {520} (\bibinfo {year}
  {2016})}\BibitemShut {NoStop}%
\bibitem [{\citenamefont {Garg}\ \emph {et~al.}(2018)\citenamefont {Garg},
  \citenamefont {Kim},\ and\ \citenamefont
  {Goulielmakis}}]{Garg_2018_NP_12_291}%
  \BibitemOpen
  \bibfield  {author} {\bibinfo {author} {\bibfnamefont {M.}~\bibnamefont
  {Garg}}, \bibinfo {author} {\bibfnamefont {H.~Y.}\ \bibnamefont {Kim}}, \
  and\ \bibinfo {author} {\bibfnamefont {E.}~\bibnamefont {Goulielmakis}},\
  }\href {\doibase 10.1038/s41566-018-0123-6} {\bibfield  {journal} {\bibinfo
  {journal} {Nat. Phot.}\ }\textbf {\bibinfo {volume} {12}},\ \bibinfo {pages}
  {291} (\bibinfo {year} {2018})}\BibitemShut {NoStop}%
\bibitem [{\citenamefont {Schubert}\ \emph {et~al.}(2014)\citenamefont
  {Schubert}, \citenamefont {Hohenleutner}, \citenamefont {Langer},
  \citenamefont {Urbanek}, \citenamefont {Lange}, \citenamefont {Huttner},
  \citenamefont {Golde}, \citenamefont {Meier}, \citenamefont {Kira},
  \citenamefont {Koch},\ and\ \citenamefont {Huber}}]{Schubert_2014_NP_8_119}%
  \BibitemOpen
  \bibfield  {author} {\bibinfo {author} {\bibfnamefont {O.}~\bibnamefont
  {Schubert}}, \bibinfo {author} {\bibfnamefont {M.}~\bibnamefont
  {Hohenleutner}}, \bibinfo {author} {\bibfnamefont {F.}~\bibnamefont
  {Langer}}, \bibinfo {author} {\bibfnamefont {B.}~\bibnamefont {Urbanek}},
  \bibinfo {author} {\bibfnamefont {C.}~\bibnamefont {Lange}}, \bibinfo
  {author} {\bibfnamefont {U.}~\bibnamefont {Huttner}}, \bibinfo {author}
  {\bibfnamefont {D.}~\bibnamefont {Golde}}, \bibinfo {author} {\bibfnamefont
  {T.}~\bibnamefont {Meier}}, \bibinfo {author} {\bibfnamefont
  {M.}~\bibnamefont {Kira}}, \bibinfo {author} {\bibfnamefont {S.~W.}\
  \bibnamefont {Koch}}, \ and\ \bibinfo {author} {\bibfnamefont
  {R.}~\bibnamefont {Huber}},\ }\href {\doibase 10.1038/nphoton.2013.349}
  {\bibfield  {journal} {\bibinfo  {journal} {Nat. Phot.}\ }\textbf {\bibinfo
  {volume} {8}},\ \bibinfo {pages} {119} (\bibinfo {year} {2014})}\BibitemShut
  {NoStop}%
\bibitem [{\citenamefont {Hohenleutner}\ \emph {et~al.}(2015)\citenamefont
  {Hohenleutner}, \citenamefont {Langer}, \citenamefont {Schubert},
  \citenamefont {Knorr}, \citenamefont {Huttner}, \citenamefont {Koch},
  \citenamefont {Kira},\ and\ \citenamefont
  {Huber}}]{Hohenleutner_2015_Nature_523_572}%
  \BibitemOpen
  \bibfield  {author} {\bibinfo {author} {\bibfnamefont {M.}~\bibnamefont
  {Hohenleutner}}, \bibinfo {author} {\bibfnamefont {F.}~\bibnamefont
  {Langer}}, \bibinfo {author} {\bibfnamefont {O.}~\bibnamefont {Schubert}},
  \bibinfo {author} {\bibfnamefont {M.}~\bibnamefont {Knorr}}, \bibinfo
  {author} {\bibfnamefont {U.}~\bibnamefont {Huttner}}, \bibinfo {author}
  {\bibfnamefont {S.~W.}\ \bibnamefont {Koch}}, \bibinfo {author}
  {\bibfnamefont {M.}~\bibnamefont {Kira}}, \ and\ \bibinfo {author}
  {\bibfnamefont {R.}~\bibnamefont {Huber}},\ }\href {\doibase
  10.1038/nature14652} {\bibfield  {journal} {\bibinfo  {journal} {Nature}\
  }\textbf {\bibinfo {volume} {523}},\ \bibinfo {pages} {572} (\bibinfo {year}
  {2015})}\BibitemShut {NoStop}%
\bibitem [{\citenamefont {Kruchinin}\ \emph {et~al.}(2013)\citenamefont
  {Kruchinin}, \citenamefont {Korbman},\ and\ \citenamefont
  {Yakovlev}}]{Kruchinin_2013_PRB_87_115201}%
  \BibitemOpen
  \bibfield  {author} {\bibinfo {author} {\bibfnamefont {S.~{\relax Yu}.}\
  \bibnamefont {Kruchinin}}, \bibinfo {author} {\bibfnamefont {M.}~\bibnamefont
  {Korbman}}, \ and\ \bibinfo {author} {\bibfnamefont {V.~S.}\ \bibnamefont
  {Yakovlev}},\ }\href {\doibase 10.1103/PhysRevB.87.115201} {\bibfield
  {journal} {\bibinfo  {journal} {Phys. Rev. B}\ }\textbf {\bibinfo {volume}
  {87}},\ \bibinfo {pages} {115201} (\bibinfo {year} {2013})}\BibitemShut
  {NoStop}%
\bibitem [{\citenamefont {Paasch-Colberg}\ \emph {et~al.}(2016)\citenamefont
  {Paasch-Colberg}, \citenamefont {Kruchinin}, \citenamefont {Sa\u{g}lam},
  \citenamefont {Kapser}, \citenamefont {Cabrini}, \citenamefont {Muehlbrandt},
  \citenamefont {Reichert}, \citenamefont {Barth}, \citenamefont {Ernstorfer},
  \citenamefont {Kienberger}, \citenamefont {Yakovlev}, \citenamefont
  {Karpowicz},\ and\ \citenamefont
  {Schiffrin}}]{Paasch-Colberg_2016_Optica_3_1358}%
  \BibitemOpen
  \bibfield  {author} {\bibinfo {author} {\bibfnamefont {T.}~\bibnamefont
  {Paasch-Colberg}}, \bibinfo {author} {\bibfnamefont {S.~{\relax Yu}.}\
  \bibnamefont {Kruchinin}}, \bibinfo {author} {\bibfnamefont
  {{\"O}.}~\bibnamefont {Sa\u{g}lam}}, \bibinfo {author} {\bibfnamefont
  {S.}~\bibnamefont {Kapser}}, \bibinfo {author} {\bibfnamefont
  {S.}~\bibnamefont {Cabrini}}, \bibinfo {author} {\bibfnamefont
  {S.}~\bibnamefont {Muehlbrandt}}, \bibinfo {author} {\bibfnamefont
  {J.}~\bibnamefont {Reichert}}, \bibinfo {author} {\bibfnamefont {J.~V.}\
  \bibnamefont {Barth}}, \bibinfo {author} {\bibfnamefont {R.}~\bibnamefont
  {Ernstorfer}}, \bibinfo {author} {\bibfnamefont {R.}~\bibnamefont
  {Kienberger}}, \bibinfo {author} {\bibfnamefont {V.~S.}\ \bibnamefont
  {Yakovlev}}, \bibinfo {author} {\bibfnamefont {N.}~\bibnamefont {Karpowicz}},
  \ and\ \bibinfo {author} {\bibfnamefont {A.}~\bibnamefont {Schiffrin}},\
  }\href {\doibase 10.1364/OPTICA.3.001358} {\bibfield  {journal} {\bibinfo
  {journal} {Optica}\ }\textbf {\bibinfo {volume} {3}},\ \bibinfo {pages}
  {1358} (\bibinfo {year} {2016})}\BibitemShut {NoStop}%
\bibitem [{\citenamefont {Schultze}\ \emph {et~al.}(2013)\citenamefont
  {Schultze}, \citenamefont {Bothschafter}, \citenamefont {Sommer},
  \citenamefont {Holzner}, \citenamefont {Schweinberger}, \citenamefont
  {Fiess}, \citenamefont {Hofstetter}, \citenamefont {Kienberger},
  \citenamefont {Apalkov}, \citenamefont {Yakovlev}, \citenamefont {Stockman},\
  and\ \citenamefont {Krausz}}]{Schultze_2013_Nature_493_75}%
  \BibitemOpen
  \bibfield  {author} {\bibinfo {author} {\bibfnamefont {M.}~\bibnamefont
  {Schultze}}, \bibinfo {author} {\bibfnamefont {E.~M.}\ \bibnamefont
  {Bothschafter}}, \bibinfo {author} {\bibfnamefont {A.}~\bibnamefont
  {Sommer}}, \bibinfo {author} {\bibfnamefont {S.}~\bibnamefont {Holzner}},
  \bibinfo {author} {\bibfnamefont {W.}~\bibnamefont {Schweinberger}}, \bibinfo
  {author} {\bibfnamefont {M.}~\bibnamefont {Fiess}}, \bibinfo {author}
  {\bibfnamefont {M.}~\bibnamefont {Hofstetter}}, \bibinfo {author}
  {\bibfnamefont {R.}~\bibnamefont {Kienberger}}, \bibinfo {author}
  {\bibfnamefont {V.}~\bibnamefont {Apalkov}}, \bibinfo {author} {\bibfnamefont
  {V.~S.}\ \bibnamefont {Yakovlev}}, \bibinfo {author} {\bibfnamefont {M.~I.}\
  \bibnamefont {Stockman}}, \ and\ \bibinfo {author} {\bibfnamefont
  {F.}~\bibnamefont {Krausz}},\ }\href {\doibase 10.1038/nature11720}
  {\bibfield  {journal} {\bibinfo  {journal} {Nature}\ }\textbf {\bibinfo
  {volume} {493}},\ \bibinfo {pages} {75} (\bibinfo {year} {2013})}\BibitemShut
  {NoStop}%
\bibitem [{\citenamefont {Schultze}\ \emph {et~al.}(2014)\citenamefont
  {Schultze}, \citenamefont {Ramasesha}, \citenamefont {Pemmaraju},
  \citenamefont {Sato}, \citenamefont {Whitmore}, \citenamefont {Gandman},
  \citenamefont {Prell}, \citenamefont {Borja}, \citenamefont {Prendergast},
  \citenamefont {Yabana}, \citenamefont {Neumark},\ and\ \citenamefont
  {Leone}}]{Schultze_2014_Sci_346_1348}%
  \BibitemOpen
  \bibfield  {author} {\bibinfo {author} {\bibfnamefont {M.}~\bibnamefont
  {Schultze}}, \bibinfo {author} {\bibfnamefont {K.}~\bibnamefont {Ramasesha}},
  \bibinfo {author} {\bibfnamefont {C.}~\bibnamefont {Pemmaraju}}, \bibinfo
  {author} {\bibfnamefont {S.}~\bibnamefont {Sato}}, \bibinfo {author}
  {\bibfnamefont {D.}~\bibnamefont {Whitmore}}, \bibinfo {author}
  {\bibfnamefont {A.}~\bibnamefont {Gandman}}, \bibinfo {author} {\bibfnamefont
  {J.~S.}\ \bibnamefont {Prell}}, \bibinfo {author} {\bibfnamefont {L.~J.}\
  \bibnamefont {Borja}}, \bibinfo {author} {\bibfnamefont {D.}~\bibnamefont
  {Prendergast}}, \bibinfo {author} {\bibfnamefont {K.}~\bibnamefont {Yabana}},
  \bibinfo {author} {\bibfnamefont {D.~M.}\ \bibnamefont {Neumark}}, \ and\
  \bibinfo {author} {\bibfnamefont {S.~R.}\ \bibnamefont {Leone}},\ }\href
  {\doibase 10.1126/science.1260311} {\bibfield  {journal} {\bibinfo  {journal}
  {Science}\ }\textbf {\bibinfo {volume} {346}},\ \bibinfo {pages} {1348}
  (\bibinfo {year} {2014})}\BibitemShut {NoStop}%
\bibitem [{\citenamefont {Lucchini}\ \emph {et~al.}(2016)\citenamefont
  {Lucchini}, \citenamefont {Sato}, \citenamefont {Ludwig}, \citenamefont
  {Herrmann}, \citenamefont {Volkov}, \citenamefont {Kasmi}, \citenamefont
  {Shinohara}, \citenamefont {Yabana}, \citenamefont {Gallmann},\ and\
  \citenamefont {Keller}}]{Lucchini_2016_Science_353_916}%
  \BibitemOpen
  \bibfield  {author} {\bibinfo {author} {\bibfnamefont {M.}~\bibnamefont
  {Lucchini}}, \bibinfo {author} {\bibfnamefont {S.~A.}\ \bibnamefont {Sato}},
  \bibinfo {author} {\bibfnamefont {A.}~\bibnamefont {Ludwig}}, \bibinfo
  {author} {\bibfnamefont {J.}~\bibnamefont {Herrmann}}, \bibinfo {author}
  {\bibfnamefont {M.}~\bibnamefont {Volkov}}, \bibinfo {author} {\bibfnamefont
  {L.}~\bibnamefont {Kasmi}}, \bibinfo {author} {\bibfnamefont
  {Y.}~\bibnamefont {Shinohara}}, \bibinfo {author} {\bibfnamefont
  {K.}~\bibnamefont {Yabana}}, \bibinfo {author} {\bibfnamefont
  {L.}~\bibnamefont {Gallmann}}, \ and\ \bibinfo {author} {\bibfnamefont
  {U.}~\bibnamefont {Keller}},\ }\href {\doibase 10.1126/science.aag1268}
  {\bibfield  {journal} {\bibinfo  {journal} {Science}\ }\textbf {\bibinfo
  {volume} {353}},\ \bibinfo {pages} {916} (\bibinfo {year}
  {2016})}\BibitemShut {NoStop}%
\bibitem [{\citenamefont {Floss}\ \emph {et~al.}(2018)\citenamefont {Floss},
  \citenamefont {Lemell}, \citenamefont {Wachter}, \citenamefont {Smejkal},
  \citenamefont {Sato}, \citenamefont {Tong}, \citenamefont {Yabana},\ and\
  \citenamefont {Burgd\"orfer}}]{Floss_2018_PRA_97_011401}%
  \BibitemOpen
  \bibfield  {author} {\bibinfo {author} {\bibfnamefont {I.}~\bibnamefont
  {Floss}}, \bibinfo {author} {\bibfnamefont {C.}~\bibnamefont {Lemell}},
  \bibinfo {author} {\bibfnamefont {G.}~\bibnamefont {Wachter}}, \bibinfo
  {author} {\bibfnamefont {V.}~\bibnamefont {Smejkal}}, \bibinfo {author}
  {\bibfnamefont {S.~A.}\ \bibnamefont {Sato}}, \bibinfo {author}
  {\bibfnamefont {X.-M.}\ \bibnamefont {Tong}}, \bibinfo {author}
  {\bibfnamefont {K.}~\bibnamefont {Yabana}}, \ and\ \bibinfo {author}
  {\bibfnamefont {J.}~\bibnamefont {Burgd\"orfer}},\ }\href {\doibase
  10.1103/PhysRevA.97.011401} {\bibfield  {journal} {\bibinfo  {journal} {Phys.
  Rev. A}\ }\textbf {\bibinfo {volume} {97}},\ \bibinfo {pages} {011401(R)}
  (\bibinfo {year} {2018})}\BibitemShut {NoStop}%
\bibitem [{\citenamefont {Otobe}(2016)}]{Otobe_2016_PRB_94_235152}%
  \BibitemOpen
  \bibfield  {author} {\bibinfo {author} {\bibfnamefont {T.}~\bibnamefont
  {Otobe}},\ }\href {\doibase 10.1103/PhysRevB.94.235152} {\bibfield  {journal}
  {\bibinfo  {journal} {Phys. Rev. B}\ }\textbf {\bibinfo {volume} {94}},\
  \bibinfo {pages} {235152} (\bibinfo {year} {2016})}\BibitemShut {NoStop}%
\bibitem [{\citenamefont {Fischetti}\ \emph {et~al.}(1985)\citenamefont
  {Fischetti}, \citenamefont {DiMaria}, \citenamefont {Brorson}, \citenamefont
  {Theis},\ and\ \citenamefont {Kirtley}}]{Fischetti_1985_PRB_31_8124}%
  \BibitemOpen
  \bibfield  {author} {\bibinfo {author} {\bibfnamefont {M.~V.}\ \bibnamefont
  {Fischetti}}, \bibinfo {author} {\bibfnamefont {D.~J.}\ \bibnamefont
  {DiMaria}}, \bibinfo {author} {\bibfnamefont {S.~D.}\ \bibnamefont
  {Brorson}}, \bibinfo {author} {\bibfnamefont {T.~N.}\ \bibnamefont {Theis}},
  \ and\ \bibinfo {author} {\bibfnamefont {J.~R.}\ \bibnamefont {Kirtley}},\
  }\href {\doibase 10.1103/PhysRevB.31.8124} {\bibfield  {journal} {\bibinfo
  {journal} {Phys. Rev. B}\ }\textbf {\bibinfo {volume} {31}},\ \bibinfo
  {pages} {8124} (\bibinfo {year} {1985})}\BibitemShut {NoStop}%
\bibitem [{\citenamefont {Arnold}\ \emph {et~al.}(1994)\citenamefont {Arnold},
  \citenamefont {Cartier},\ and\ \citenamefont
  {DiMaria}}]{Arnold_1994_PRB_49_10278}%
  \BibitemOpen
  \bibfield  {author} {\bibinfo {author} {\bibfnamefont {D.}~\bibnamefont
  {Arnold}}, \bibinfo {author} {\bibfnamefont {E.}~\bibnamefont {Cartier}}, \
  and\ \bibinfo {author} {\bibfnamefont {D.~J.}\ \bibnamefont {DiMaria}},\
  }\href {\doibase 10.1103/PhysRevB.49.10278} {\bibfield  {journal} {\bibinfo
  {journal} {Phys. Rev. B}\ }\textbf {\bibinfo {volume} {49}},\ \bibinfo
  {pages} {10278} (\bibinfo {year} {1994})}\BibitemShut {NoStop}%
\bibitem [{\citenamefont {Kira}\ and\ \citenamefont
  {Koch}(2006)}]{Kira_2006_PQE_30_155}%
  \BibitemOpen
  \bibfield  {author} {\bibinfo {author} {\bibfnamefont {M.}~\bibnamefont
  {Kira}}\ and\ \bibinfo {author} {\bibfnamefont {S.}~\bibnamefont {Koch}},\
  }\href {\doibase https://doi.org/10.1016/j.pquantelec.2006.12.002} {\bibfield
   {journal} {\bibinfo  {journal} {Prog. Quant. Electr.}\ }\textbf {\bibinfo
  {volume} {30}},\ \bibinfo {pages} {155 } (\bibinfo {year}
  {2006})}\BibitemShut {NoStop}%
\bibitem [{\citenamefont {Smith}\ \emph {et~al.}(2010)\citenamefont {Smith},
  \citenamefont {Wahlstrand}, \citenamefont {Funk}, \citenamefont {Mirin},
  \citenamefont {Cundiff}, \citenamefont {Steiner}, \citenamefont {Schafer},
  \citenamefont {Kira},\ and\ \citenamefont
  {Koch}}]{Smith_2010_PRL_104_247401}%
  \BibitemOpen
  \bibfield  {author} {\bibinfo {author} {\bibfnamefont {R.~P.}\ \bibnamefont
  {Smith}}, \bibinfo {author} {\bibfnamefont {J.~K.}\ \bibnamefont
  {Wahlstrand}}, \bibinfo {author} {\bibfnamefont {A.~C.}\ \bibnamefont
  {Funk}}, \bibinfo {author} {\bibfnamefont {R.~P.}\ \bibnamefont {Mirin}},
  \bibinfo {author} {\bibfnamefont {S.~T.}\ \bibnamefont {Cundiff}}, \bibinfo
  {author} {\bibfnamefont {J.~T.}\ \bibnamefont {Steiner}}, \bibinfo {author}
  {\bibfnamefont {M.}~\bibnamefont {Schafer}}, \bibinfo {author} {\bibfnamefont
  {M.}~\bibnamefont {Kira}}, \ and\ \bibinfo {author} {\bibfnamefont {S.~W.}\
  \bibnamefont {Koch}},\ }\href {\doibase 10.1103/PhysRevLett.104.247401}
  {\bibfield  {journal} {\bibinfo  {journal} {Phys. Rev. Lett.}\ }\textbf
  {\bibinfo {volume} {104}},\ \bibinfo {pages} {247401} (\bibinfo {year}
  {2010})}\BibitemShut {NoStop}%
\bibitem [{\citenamefont {Vampa}\ \emph {et~al.}(2015)\citenamefont {Vampa},
  \citenamefont {Hammond}, \citenamefont {Thire}, \citenamefont {Schmidt},
  \citenamefont {Legare}, \citenamefont {McDonald}, \citenamefont {Brabec},
  \citenamefont {Klug},\ and\ \citenamefont
  {Corkum}}]{Vampa_2015_PRL_115_193603}%
  \BibitemOpen
  \bibfield  {author} {\bibinfo {author} {\bibfnamefont {G.}~\bibnamefont
  {Vampa}}, \bibinfo {author} {\bibfnamefont {T.~J.}\ \bibnamefont {Hammond}},
  \bibinfo {author} {\bibfnamefont {N.}~\bibnamefont {Thire}}, \bibinfo
  {author} {\bibfnamefont {B.~E.}\ \bibnamefont {Schmidt}}, \bibinfo {author}
  {\bibfnamefont {F.}~\bibnamefont {Legare}}, \bibinfo {author} {\bibfnamefont
  {C.~R.}\ \bibnamefont {McDonald}}, \bibinfo {author} {\bibfnamefont
  {T.}~\bibnamefont {Brabec}}, \bibinfo {author} {\bibfnamefont {D.~D.}\
  \bibnamefont {Klug}}, \ and\ \bibinfo {author} {\bibfnamefont {P.~B.}\
  \bibnamefont {Corkum}},\ }\href {\doibase 10.1103/PhysRevLett.115.193603}
  {\bibfield  {journal} {\bibinfo  {journal} {Phys. Rev. Lett.}\ }\textbf
  {\bibinfo {volume} {115}},\ \bibinfo {pages} {193603} (\bibinfo {year}
  {2015})}\BibitemShut {NoStop}%
\bibitem [{\citenamefont {Tokuyama}\ and\ \citenamefont
  {Mori}(1976)}]{Tokuyama_1976_PTP_55_411}%
  \BibitemOpen
  \bibfield  {author} {\bibinfo {author} {\bibfnamefont {M.}~\bibnamefont
  {Tokuyama}}\ and\ \bibinfo {author} {\bibfnamefont {H.}~\bibnamefont
  {Mori}},\ }\href {\doibase 10.1143/PTP.55.411} {\bibfield  {journal}
  {\bibinfo  {journal} {Prog. Theor. Phys.}\ }\textbf {\bibinfo {volume}
  {55}},\ \bibinfo {pages} {411} (\bibinfo {year} {1976})}\BibitemShut
  {NoStop}%
\bibitem [{\citenamefont {Ahn}(1994)}]{Ahn_1994_PRB_50_8310}%
  \BibitemOpen
  \bibfield  {author} {\bibinfo {author} {\bibfnamefont {D.}~\bibnamefont
  {Ahn}},\ }\href {\doibase 10.1103/PhysRevB.50.8310} {\bibfield  {journal}
  {\bibinfo  {journal} {Phys. Rev. B}\ }\textbf {\bibinfo {volume} {50}},\
  \bibinfo {pages} {8310} (\bibinfo {year} {1994})}\BibitemShut {NoStop}%
\bibitem [{\citenamefont {Breuer}\ \emph {et~al.}(2016)\citenamefont {Breuer},
  \citenamefont {Laine}, \citenamefont {Piilo},\ and\ \citenamefont
  {Vacchini}}]{Breuer_2016_RMP_88_021002}%
  \BibitemOpen
  \bibfield  {author} {\bibinfo {author} {\bibfnamefont {H.-P.}\ \bibnamefont
  {Breuer}}, \bibinfo {author} {\bibfnamefont {E.-M.}\ \bibnamefont {Laine}},
  \bibinfo {author} {\bibfnamefont {J.}~\bibnamefont {Piilo}}, \ and\ \bibinfo
  {author} {\bibfnamefont {B.}~\bibnamefont {Vacchini}},\ }\href {\doibase
  10.1103/RevModPhys.88.021002} {\bibfield  {journal} {\bibinfo  {journal}
  {Rev. Mod. Phys.}\ }\textbf {\bibinfo {volume} {88}},\ \bibinfo {pages}
  {021002} (\bibinfo {year} {2016})}\BibitemShut {NoStop}%
\bibitem [{\citenamefont {de~Vega}\ and\ \citenamefont
  {Alonso}(2017)}]{deVega_2017_RMP_89_015001}%
  \BibitemOpen
  \bibfield  {author} {\bibinfo {author} {\bibfnamefont {I.}~\bibnamefont
  {de~Vega}}\ and\ \bibinfo {author} {\bibfnamefont {D.}~\bibnamefont
  {Alonso}},\ }\href {\doibase 10.1103/RevModPhys.89.015001} {\bibfield
  {journal} {\bibinfo  {journal} {Rev. Mod. Phys.}\ }\textbf {\bibinfo {volume}
  {89}},\ \bibinfo {pages} {015001} (\bibinfo {year} {2017})}\BibitemShut
  {NoStop}%
\bibitem [{\citenamefont {Nam}\ \emph {et~al.}(1976)\citenamefont {Nam},
  \citenamefont {Reynolds}, \citenamefont {Litton}, \citenamefont {Almassy},
  \citenamefont {Collins},\ and\ \citenamefont {Wolfe}}]{Nam_1976_PRB_13_761}%
  \BibitemOpen
  \bibfield  {author} {\bibinfo {author} {\bibfnamefont {S.~B.}\ \bibnamefont
  {Nam}}, \bibinfo {author} {\bibfnamefont {D.~C.}\ \bibnamefont {Reynolds}},
  \bibinfo {author} {\bibfnamefont {C.~W.}\ \bibnamefont {Litton}}, \bibinfo
  {author} {\bibfnamefont {R.~J.}\ \bibnamefont {Almassy}}, \bibinfo {author}
  {\bibfnamefont {T.~C.}\ \bibnamefont {Collins}}, \ and\ \bibinfo {author}
  {\bibfnamefont {C.~M.}\ \bibnamefont {Wolfe}},\ }\href {\doibase
  10.1103/PhysRevB.13.761} {\bibfield  {journal} {\bibinfo  {journal} {Phys.
  Rev. B}\ }\textbf {\bibinfo {volume} {13}},\ \bibinfo {pages} {761} (\bibinfo
  {year} {1976})}\BibitemShut {NoStop}%
\bibitem [{\citenamefont {Green}(2013)}]{Green_2013_AIPA_3_112104}%
  \BibitemOpen
  \bibfield  {author} {\bibinfo {author} {\bibfnamefont {M.~A.}\ \bibnamefont
  {Green}},\ }\href {\doibase 10.1063/1.4828730} {\bibfield  {journal}
  {\bibinfo  {journal} {AIP Advances}\ }\textbf {\bibinfo {volume} {3}},\
  \bibinfo {pages} {112104} (\bibinfo {year} {2013})}\BibitemShut {NoStop}%
\bibitem [{\citenamefont {Kruchinin}\ \emph {et~al.}(2018)\citenamefont
  {Kruchinin}, \citenamefont {Krausz},\ and\ \citenamefont
  {Yakovlev}}]{Kruchinin_2018_RMP_90_021002}%
  \BibitemOpen
  \bibfield  {author} {\bibinfo {author} {\bibfnamefont {S.~{\relax Yu}.}\
  \bibnamefont {Kruchinin}}, \bibinfo {author} {\bibfnamefont {F.}~\bibnamefont
  {Krausz}}, \ and\ \bibinfo {author} {\bibfnamefont {V.~S.}\ \bibnamefont
  {Yakovlev}},\ }\href {\doibase 10.1103/RevModPhys.90.021002} {\bibfield
  {journal} {\bibinfo  {journal} {Rev. Mod. Phys.}\ }\textbf {\bibinfo {volume}
  {90}},\ \bibinfo {pages} {021002} (\bibinfo {year} {2018})}\BibitemShut
  {NoStop}%
\bibitem [{\citenamefont {Tsujibayashi}\ and\ \citenamefont
  {Toyoda}(2002)}]{Tsujibayashi_2002_REDS_157_969}%
  \BibitemOpen
  \bibfield  {author} {\bibinfo {author} {\bibfnamefont {T.}~\bibnamefont
  {Tsujibayashi}}\ and\ \bibinfo {author} {\bibfnamefont {K.}~\bibnamefont
  {Toyoda}},\ }\href {\doibase 10.1080/10420150215766} {\bibfield  {journal}
  {\bibinfo  {journal} {Radiation Effects and Defects in Solids}\ }\textbf
  {\bibinfo {volume} {157}},\ \bibinfo {pages} {969} (\bibinfo {year}
  {2002})}\BibitemShut {NoStop}%
\bibitem [{\citenamefont {Sugiura}(1992)}]{Sugiura_2016_JJAP_31_2816}%
  \BibitemOpen
  \bibfield  {author} {\bibinfo {author} {\bibfnamefont {C.}~\bibnamefont
  {Sugiura}},\ }\href {http://stacks.iop.org/1347-4065/31/i=9R/a=2816}
  {\bibfield  {journal} {\bibinfo  {journal} {Japanese Journal of Applied
  Physics}\ }\textbf {\bibinfo {volume} {31}},\ \bibinfo {pages} {2816}
  (\bibinfo {year} {1992})}\BibitemShut {NoStop}%
\bibitem [{\citenamefont {Kresse}\ \emph {et~al.}(2012)\citenamefont {Kresse},
  \citenamefont {Marsman}, \citenamefont {Hintzsche},\ and\ \citenamefont
  {Flage-Larsen}}]{Kresse_2012_PRB_85_045205}%
  \BibitemOpen
  \bibfield  {author} {\bibinfo {author} {\bibfnamefont {G.}~\bibnamefont
  {Kresse}}, \bibinfo {author} {\bibfnamefont {M.}~\bibnamefont {Marsman}},
  \bibinfo {author} {\bibfnamefont {L.~E.}\ \bibnamefont {Hintzsche}}, \ and\
  \bibinfo {author} {\bibfnamefont {E.}~\bibnamefont {Flage-Larsen}},\ }\href
  {\doibase 10.1103/PhysRevB.85.045205} {\bibfield  {journal} {\bibinfo
  {journal} {Phys. Rev. B}\ }\textbf {\bibinfo {volume} {85}},\ \bibinfo
  {pages} {045205} (\bibinfo {year} {2012})}\BibitemShut {NoStop}%
\bibitem [{\citenamefont {Sander}\ \emph {et~al.}(2015)\citenamefont {Sander},
  \citenamefont {Maggio},\ and\ \citenamefont
  {Kresse}}]{Sander_2015_PRB_92_045209}%
  \BibitemOpen
  \bibfield  {author} {\bibinfo {author} {\bibfnamefont {T.}~\bibnamefont
  {Sander}}, \bibinfo {author} {\bibfnamefont {E.}~\bibnamefont {Maggio}}, \
  and\ \bibinfo {author} {\bibfnamefont {G.}~\bibnamefont {Kresse}},\ }\href
  {\doibase 10.1103/PhysRevB.92.045209} {\bibfield  {journal} {\bibinfo
  {journal} {Phys. Rev. B}\ }\textbf {\bibinfo {volume} {92}},\ \bibinfo
  {pages} {045209} (\bibinfo {year} {2015})}\BibitemShut {NoStop}%
\bibitem [{\citenamefont {Mostofi}\ \emph {et~al.}(2014)\citenamefont
  {Mostofi}, \citenamefont {Yates}, \citenamefont {Pizzi}, \citenamefont {Lee},
  \citenamefont {Souza}, \citenamefont {Vanderbilt},\ and\ \citenamefont
  {Marzari}}]{Mostofi_2014_CPC_185_2309}%
  \BibitemOpen
  \bibfield  {author} {\bibinfo {author} {\bibfnamefont {A.~A.}\ \bibnamefont
  {Mostofi}}, \bibinfo {author} {\bibfnamefont {J.~R.}\ \bibnamefont {Yates}},
  \bibinfo {author} {\bibfnamefont {G.}~\bibnamefont {Pizzi}}, \bibinfo
  {author} {\bibfnamefont {Y.-S.}\ \bibnamefont {Lee}}, \bibinfo {author}
  {\bibfnamefont {I.}~\bibnamefont {Souza}}, \bibinfo {author} {\bibfnamefont
  {D.}~\bibnamefont {Vanderbilt}}, \ and\ \bibinfo {author} {\bibfnamefont
  {N.}~\bibnamefont {Marzari}},\ }\href {\doibase 10.1016/j.cpc.2014.05.003}
  {\bibfield  {journal} {\bibinfo  {journal} {Comp. Phys. Comm.}\ }\textbf
  {\bibinfo {volume} {185}},\ \bibinfo {pages} {2309} (\bibinfo {year}
  {2014})}\BibitemShut {NoStop}%
\bibitem [{\citenamefont {Bloch}(1928)}]{Bloch_1928_ZP_52_555}%
  \BibitemOpen
  \bibfield  {author} {\bibinfo {author} {\bibfnamefont {F.}~\bibnamefont
  {Bloch}},\ }\href {\doibase 10.1007/BF01339455} {\bibfield  {journal}
  {\bibinfo  {journal} {Z. Physik}\ }\textbf {\bibinfo {volume} {52}},\
  \bibinfo {pages} {555} (\bibinfo {year} {1928})}\BibitemShut {NoStop}%
\bibitem [{\citenamefont {Wismer}\ \emph {et~al.}(2016)\citenamefont {Wismer},
  \citenamefont {Kruchinin}, \citenamefont {Ciappina}, \citenamefont
  {Stockman},\ and\ \citenamefont {Yakovlev}}]{Wismer_2016_PRL_116_197401}%
  \BibitemOpen
  \bibfield  {author} {\bibinfo {author} {\bibfnamefont {M.~S.}\ \bibnamefont
  {Wismer}}, \bibinfo {author} {\bibfnamefont {S.~{\relax Yu}.}\ \bibnamefont
  {Kruchinin}}, \bibinfo {author} {\bibfnamefont {M.}~\bibnamefont {Ciappina}},
  \bibinfo {author} {\bibfnamefont {M.~I.}\ \bibnamefont {Stockman}}, \ and\
  \bibinfo {author} {\bibfnamefont {V.~S.}\ \bibnamefont {Yakovlev}},\ }\href
  {\doibase 10.1103/PhysRevLett.116.197401} {\bibfield  {journal} {\bibinfo
  {journal} {Phys. Rev. Lett.}\ }\textbf {\bibinfo {volume} {116}},\ \bibinfo
  {pages} {197401} (\bibinfo {year} {2016})}\BibitemShut {NoStop}%
\bibitem [{\citenamefont {Yamaguchi}\ \emph {et~al.}(2017)\citenamefont
  {Yamaguchi}, \citenamefont {Yuge},\ and\ \citenamefont
  {Ogawa}}]{Yamaguchi_2017_PRE_95_012136}%
  \BibitemOpen
  \bibfield  {author} {\bibinfo {author} {\bibfnamefont {M.}~\bibnamefont
  {Yamaguchi}}, \bibinfo {author} {\bibfnamefont {T.}~\bibnamefont {Yuge}}, \
  and\ \bibinfo {author} {\bibfnamefont {T.}~\bibnamefont {Ogawa}},\ }\href
  {\doibase 10.1103/PhysRevE.95.012136} {\bibfield  {journal} {\bibinfo
  {journal} {Phys. Rev. E}\ }\textbf {\bibinfo {volume} {95}},\ \bibinfo
  {pages} {012136} (\bibinfo {year} {2017})}\BibitemShut {NoStop}%
\bibitem [{\citenamefont {Huttner}\ \emph {et~al.}(2017)\citenamefont
  {Huttner}, \citenamefont {Kira},\ and\ \citenamefont
  {Koch}}]{Huttner_2017_LPR_11_1700049}%
  \BibitemOpen
  \bibfield  {author} {\bibinfo {author} {\bibfnamefont {U.}~\bibnamefont
  {Huttner}}, \bibinfo {author} {\bibfnamefont {M.}~\bibnamefont {Kira}}, \
  and\ \bibinfo {author} {\bibfnamefont {S.~W.}\ \bibnamefont {Koch}},\ }\href
  {\doibase 10.1002/lpor.201700049} {\bibfield  {journal} {\bibinfo  {journal}
  {Laser {\&} Photonics Reviews}\ }\textbf {\bibinfo {volume} {11}},\ \bibinfo
  {pages} {1700049} (\bibinfo {year} {2017})}\BibitemShut {NoStop}%
\bibitem [{\citenamefont {Courant}\ and\ \citenamefont
  {Hilbert}(1989)}]{Courant_1989}%
  \BibitemOpen
  \bibfield  {author} {\bibinfo {author} {\bibfnamefont {R.}~\bibnamefont
  {Courant}}\ and\ \bibinfo {author} {\bibfnamefont {D.}~\bibnamefont
  {Hilbert}},\ }\href@noop {} {\emph {\bibinfo {title} {Methods of Mathematical
  Physics: Partial Differential Equations}}},\ Wiley Classics Library\
  (\bibinfo  {publisher} {Wiley},\ \bibinfo {year} {1989})\ pp.\ \bibinfo
  {pages} {62--64}\BibitemShut {NoStop}%
\bibitem [{\citenamefont {Dunlap}\ and\ \citenamefont
  {Kenkre}(1986)}]{Dunlap_1986_PRB_34_3625}%
  \BibitemOpen
  \bibfield  {author} {\bibinfo {author} {\bibfnamefont {D.~H.}\ \bibnamefont
  {Dunlap}}\ and\ \bibinfo {author} {\bibfnamefont {V.~M.}\ \bibnamefont
  {Kenkre}},\ }\href {\doibase 10.1103/PhysRevB.34.3625} {\bibfield  {journal}
  {\bibinfo  {journal} {Phys. Rev. B}\ }\textbf {\bibinfo {volume} {34}},\
  \bibinfo {pages} {3625} (\bibinfo {year} {1986})}\BibitemShut {NoStop}%
\bibitem [{\citenamefont {Krieger}\ and\ \citenamefont
  {Iafrate}(1986)}]{Krieger_1986_PRB_33_5494}%
  \BibitemOpen
  \bibfield  {author} {\bibinfo {author} {\bibfnamefont {J.~B.}\ \bibnamefont
  {Krieger}}\ and\ \bibinfo {author} {\bibfnamefont {G.~J.}\ \bibnamefont
  {Iafrate}},\ }\href {\doibase 10.1103/PhysRevB.33.5494} {\bibfield  {journal}
  {\bibinfo  {journal} {Phys. Rev. B}\ }\textbf {\bibinfo {volume} {33}},\
  \bibinfo {pages} {5494} (\bibinfo {year} {1986})}\BibitemShut {NoStop}%
\bibitem [{\citenamefont {McDonald}\ \emph {et~al.}(2015)\citenamefont
  {McDonald}, \citenamefont {Vampa}, \citenamefont {Corkum},\ and\
  \citenamefont {Brabec}}]{McDonald_2015_PRA_92_033845}%
  \BibitemOpen
  \bibfield  {author} {\bibinfo {author} {\bibfnamefont {C.~R.}\ \bibnamefont
  {McDonald}}, \bibinfo {author} {\bibfnamefont {G.}~\bibnamefont {Vampa}},
  \bibinfo {author} {\bibfnamefont {P.~B.}\ \bibnamefont {Corkum}}, \ and\
  \bibinfo {author} {\bibfnamefont {T.}~\bibnamefont {Brabec}},\ }\href
  {\doibase 10.1103/PhysRevA.92.033845} {\bibfield  {journal} {\bibinfo
  {journal} {Phys. Rev. A}\ }\textbf {\bibinfo {volume} {92}},\ \bibinfo
  {pages} {033845} (\bibinfo {year} {2015})}\BibitemShut {NoStop}%
\bibitem [{\citenamefont {Haug}\ and\ \citenamefont
  {Koch}(2009)}]{HaugKoch_2009}%
  \BibitemOpen
  \bibfield  {author} {\bibinfo {author} {\bibfnamefont {H.}~\bibnamefont
  {Haug}}\ and\ \bibinfo {author} {\bibfnamefont {S.~W.}\ \bibnamefont
  {Koch}},\ }\href {\doibase 10.1142/9789812838858} {\emph {\bibinfo {title}
  {Quantum theory of the optical and electronic properties of
  semiconductors}}}\ (\bibinfo  {publisher} {World Scientific},\ \bibinfo
  {year} {2009})\BibitemShut {NoStop}%
\bibitem [{\citenamefont {Lindberg}\ and\ \citenamefont
  {Koch}(1988)}]{Lindberg_1988_PRB_38_3342}%
  \BibitemOpen
  \bibfield  {author} {\bibinfo {author} {\bibfnamefont {M.}~\bibnamefont
  {Lindberg}}\ and\ \bibinfo {author} {\bibfnamefont {S.~W.}\ \bibnamefont
  {Koch}},\ }\href {\doibase 10.1103/PhysRevB.38.3342} {\bibfield  {journal}
  {\bibinfo  {journal} {Phys. Rev. B}\ }\textbf {\bibinfo {volume} {38}},\
  \bibinfo {pages} {3342} (\bibinfo {year} {1988})}\BibitemShut {NoStop}%
\bibitem [{\citenamefont {Sato}\ \emph {et~al.}(2018)\citenamefont {Sato},
  \citenamefont {Pathak}, \citenamefont {Orimo},\ and\ \citenamefont
  {Ishikawa}}]{Sato_2018_JCP_148_051101}%
  \BibitemOpen
  \bibfield  {author} {\bibinfo {author} {\bibfnamefont {T.}~\bibnamefont
  {Sato}}, \bibinfo {author} {\bibfnamefont {H.}~\bibnamefont {Pathak}},
  \bibinfo {author} {\bibfnamefont {Y.}~\bibnamefont {Orimo}}, \ and\ \bibinfo
  {author} {\bibfnamefont {K.~L.}\ \bibnamefont {Ishikawa}},\ }\href {\doibase
  10.1063/1.5020633} {\bibfield  {journal} {\bibinfo  {journal} {J. Chem.
  Phys.}\ }\textbf {\bibinfo {volume} {148}},\ \bibinfo {pages} {051101}
  (\bibinfo {year} {2018})}\BibitemShut {NoStop}%
\bibitem [{\citenamefont {Feynman}\ and\ \citenamefont
  {Vernon}(2000)}]{Feynman_2000_AP_281_547}%
  \BibitemOpen
  \bibfield  {author} {\bibinfo {author} {\bibfnamefont {R.}~\bibnamefont
  {Feynman}}\ and\ \bibinfo {author} {\bibfnamefont {F.}~\bibnamefont
  {Vernon}},\ }\href {\doibase 10.1006/aphy.2000.6017} {\bibfield  {journal}
  {\bibinfo  {journal} {Ann. Phys.}\ }\textbf {\bibinfo {volume} {281}},\
  \bibinfo {pages} {547 } (\bibinfo {year} {2000})}\BibitemShut {NoStop}%
\bibitem [{\citenamefont {Caldeira}\ and\ \citenamefont
  {Leggett}(1983)}]{Caldeira_1983_AP_149_374}%
  \BibitemOpen
  \bibfield  {author} {\bibinfo {author} {\bibfnamefont {A.}~\bibnamefont
  {Caldeira}}\ and\ \bibinfo {author} {\bibfnamefont {A.}~\bibnamefont
  {Leggett}},\ }\href {\doibase 10.1016/0003-4916(83)90202-6} {\bibfield
  {journal} {\bibinfo  {journal} {Annals of Physics}\ }\textbf {\bibinfo
  {volume} {149}},\ \bibinfo {pages} {374 } (\bibinfo {year}
  {1983})}\BibitemShut {NoStop}%
\bibitem [{\citenamefont {Schlosshauer}(2007)}]{Schlosshauer_2007}%
  \BibitemOpen
  \bibfield  {author} {\bibinfo {author} {\bibfnamefont {M.}~\bibnamefont
  {Schlosshauer}},\ }\href {https://books.google.at/books?id=1qrJUS5zNbEC}
  {\emph {\bibinfo {title} {Decoherence and the Quantum-To-Classical
  Transition}}},\ The Frontiers Collection\ (\bibinfo  {publisher} {Springer},\
  \bibinfo {year} {2007})\BibitemShut {NoStop}%
\bibitem [{\citenamefont {Becker}\ \emph {et~al.}(1988)\citenamefont {Becker},
  \citenamefont {Fragnito}, \citenamefont {Cruz}, \citenamefont {Fork},
  \citenamefont {Cunningham}, \citenamefont {Henry},\ and\ \citenamefont
  {Shank}}]{Becker_1988_PRL_61_1647}%
  \BibitemOpen
  \bibfield  {author} {\bibinfo {author} {\bibfnamefont {P.~C.}\ \bibnamefont
  {Becker}}, \bibinfo {author} {\bibfnamefont {H.~L.}\ \bibnamefont
  {Fragnito}}, \bibinfo {author} {\bibfnamefont {C.~H.~B.}\ \bibnamefont
  {Cruz}}, \bibinfo {author} {\bibfnamefont {R.~L.}\ \bibnamefont {Fork}},
  \bibinfo {author} {\bibfnamefont {J.~E.}\ \bibnamefont {Cunningham}},
  \bibinfo {author} {\bibfnamefont {J.~E.}\ \bibnamefont {Henry}}, \ and\
  \bibinfo {author} {\bibfnamefont {C.~V.}\ \bibnamefont {Shank}},\ }\href
  {\doibase 10.1103/PhysRevLett.61.1647} {\bibfield  {journal} {\bibinfo
  {journal} {Phys. Rev. Lett.}\ }\textbf {\bibinfo {volume} {61}},\ \bibinfo
  {pages} {1647} (\bibinfo {year} {1988})}\BibitemShut {NoStop}%
\bibitem [{\citenamefont {Wang}\ \emph {et~al.}(1993)\citenamefont {Wang},
  \citenamefont {Ferrio}, \citenamefont {Steel}, \citenamefont {Hu},
  \citenamefont {Binder},\ and\ \citenamefont {Koch}}]{Wang_1993_PRL_71_1261}%
  \BibitemOpen
  \bibfield  {author} {\bibinfo {author} {\bibfnamefont {H.}~\bibnamefont
  {Wang}}, \bibinfo {author} {\bibfnamefont {K.}~\bibnamefont {Ferrio}},
  \bibinfo {author} {\bibfnamefont {D.~G.}\ \bibnamefont {Steel}}, \bibinfo
  {author} {\bibfnamefont {Y.~Z.}\ \bibnamefont {Hu}}, \bibinfo {author}
  {\bibfnamefont {R.}~\bibnamefont {Binder}}, \ and\ \bibinfo {author}
  {\bibfnamefont {S.~W.}\ \bibnamefont {Koch}},\ }\href {\doibase
  10.1103/PhysRevLett.71.1261} {\bibfield  {journal} {\bibinfo  {journal}
  {Phys. Rev. Lett.}\ }\textbf {\bibinfo {volume} {71}},\ \bibinfo {pages}
  {1261} (\bibinfo {year} {1993})}\BibitemShut {NoStop}%
\bibitem [{\citenamefont {H\"ugel}\ \emph {et~al.}(1999)\citenamefont
  {H\"ugel}, \citenamefont {Heinrich}, \citenamefont {Wegener}, \citenamefont
  {Vu}, \citenamefont {B\'anyai},\ and\ \citenamefont
  {Haug}}]{Huegel_1999_PRL_83_3313}%
  \BibitemOpen
  \bibfield  {author} {\bibinfo {author} {\bibfnamefont {W.~A.}\ \bibnamefont
  {H\"ugel}}, \bibinfo {author} {\bibfnamefont {M.~F.}\ \bibnamefont
  {Heinrich}}, \bibinfo {author} {\bibfnamefont {M.}~\bibnamefont {Wegener}},
  \bibinfo {author} {\bibfnamefont {Q.~T.}\ \bibnamefont {Vu}}, \bibinfo
  {author} {\bibfnamefont {L.}~\bibnamefont {B\'anyai}}, \ and\ \bibinfo
  {author} {\bibfnamefont {H.}~\bibnamefont {Haug}},\ }\href {\doibase
  10.1103/PhysRevLett.83.3313} {\bibfield  {journal} {\bibinfo  {journal}
  {Phys. Rev. Lett.}\ }\textbf {\bibinfo {volume} {83}},\ \bibinfo {pages}
  {3313} (\bibinfo {year} {1999})}\BibitemShut {NoStop}%
\bibitem [{\citenamefont {Bechstedt}(2015)}]{Bechstedt_2015}%
  \BibitemOpen
  \bibfield  {author} {\bibinfo {author} {\bibfnamefont {F.}~\bibnamefont
  {Bechstedt}},\ }\href {\doibase 10.1007/978-3-662-44593-8} {\emph {\bibinfo
  {title} {Many-Body Approach to Electronic Excitations}}}\ (\bibinfo
  {publisher} {Springer},\ \bibinfo {year} {2015})\ p.~\bibinfo {pages}
  {8}\BibitemShut {NoStop}%
\bibitem [{\citenamefont {Born}\ and\ \citenamefont
  {Wolf}(1980)}]{BornWolf_1980}%
  \BibitemOpen
  \bibfield  {author} {\bibinfo {author} {\bibfnamefont {M.}~\bibnamefont
  {Born}}\ and\ \bibinfo {author} {\bibfnamefont {E.}~\bibnamefont {Wolf}},\
  }\href {https://books.google.at/books?id=DBgIAQAAIAAJ} {\emph {\bibinfo
  {title} {Principles of Optics: Electromagnetic Theory of Propagation,
  Interference and Diffraction of Light}}}\ (\bibinfo  {publisher} {Elsevier
  Science Limited},\ \bibinfo {year} {1980})\BibitemShut {NoStop}%
\bibitem [{\citenamefont {Hawkins}\ and\ \citenamefont
  {Ivanov}(2013)}]{Hawkins_2013_PRA_87_063842}%
  \BibitemOpen
  \bibfield  {author} {\bibinfo {author} {\bibfnamefont {P.~G.}\ \bibnamefont
  {Hawkins}}\ and\ \bibinfo {author} {\bibfnamefont {M.~{\relax Yu}.}\
  \bibnamefont {Ivanov}},\ }\href {\doibase 10.1103/PhysRevA.87.063842}
  {\bibfield  {journal} {\bibinfo  {journal} {Phys. Rev. A}\ }\textbf {\bibinfo
  {volume} {87}},\ \bibinfo {pages} {063842} (\bibinfo {year}
  {2013})}\BibitemShut {NoStop}%
\bibitem [{\citenamefont {Zhokhov}\ and\ \citenamefont
  {Zheltikov}(2014)}]{Zhokhov_2014_PRL_113_133903}%
  \BibitemOpen
  \bibfield  {author} {\bibinfo {author} {\bibfnamefont {P.~A.}\ \bibnamefont
  {Zhokhov}}\ and\ \bibinfo {author} {\bibfnamefont {A.~M.}\ \bibnamefont
  {Zheltikov}},\ }\href {\doibase 10.1103/PhysRevLett.113.133903} {\bibfield
  {journal} {\bibinfo  {journal} {Phys. Rev. Lett.}\ }\textbf {\bibinfo
  {volume} {113}},\ \bibinfo {pages} {133903} (\bibinfo {year}
  {2014})}\BibitemShut {NoStop}%
\bibitem [{\citenamefont {Shcheblanov}\ \emph {et~al.}(2017)\citenamefont
  {Shcheblanov}, \citenamefont {Povarnitsyn}, \citenamefont {Terekhin},
  \citenamefont {Guizard},\ and\ \citenamefont
  {Couairon}}]{Shcheblanov_2017_PRA_96_063410}%
  \BibitemOpen
  \bibfield  {author} {\bibinfo {author} {\bibfnamefont {N.~S.}\ \bibnamefont
  {Shcheblanov}}, \bibinfo {author} {\bibfnamefont {M.~E.}\ \bibnamefont
  {Povarnitsyn}}, \bibinfo {author} {\bibfnamefont {P.~N.}\ \bibnamefont
  {Terekhin}}, \bibinfo {author} {\bibfnamefont {S.}~\bibnamefont {Guizard}}, \
  and\ \bibinfo {author} {\bibfnamefont {A.}~\bibnamefont {Couairon}},\ }\href
  {\doibase 10.1103/PhysRevA.96.063410} {\bibfield  {journal} {\bibinfo
  {journal} {Phys. Rev. A}\ }\textbf {\bibinfo {volume} {96}},\ \bibinfo
  {pages} {063410} (\bibinfo {year} {2017})}\BibitemShut {NoStop}%
\bibitem [{\citenamefont {Schuh}\ \emph {et~al.}(2017)\citenamefont {Schuh},
  \citenamefont {Kolesik}, \citenamefont {Wright}, \citenamefont {Moloney},\
  and\ \citenamefont {Koch}}]{Schuh_2017_PRL_118_063901}%
  \BibitemOpen
  \bibfield  {author} {\bibinfo {author} {\bibfnamefont {K.}~\bibnamefont
  {Schuh}}, \bibinfo {author} {\bibfnamefont {M.}~\bibnamefont {Kolesik}},
  \bibinfo {author} {\bibfnamefont {E.~M.}\ \bibnamefont {Wright}}, \bibinfo
  {author} {\bibfnamefont {J.~V.}\ \bibnamefont {Moloney}}, \ and\ \bibinfo
  {author} {\bibfnamefont {S.~W.}\ \bibnamefont {Koch}},\ }\href {\doibase
  10.1103/PhysRevLett.118.063901} {\bibfield  {journal} {\bibinfo  {journal}
  {Phys. Rev. Lett.}\ }\textbf {\bibinfo {volume} {118}},\ \bibinfo {pages}
  {063901} (\bibinfo {year} {2017})}\BibitemShut {NoStop}%
\bibitem [{\citenamefont {Chen}\ \emph {et~al.}(2018)\citenamefont {Chen},
  \citenamefont {Zhang}, \citenamefont {Chen},\ and\ \citenamefont
  {Franco}}]{Chen_2018_NC_9_2070}%
  \BibitemOpen
  \bibfield  {author} {\bibinfo {author} {\bibfnamefont {L.}~\bibnamefont
  {Chen}}, \bibinfo {author} {\bibfnamefont {Y.}~\bibnamefont {Zhang}},
  \bibinfo {author} {\bibfnamefont {G.}~\bibnamefont {Chen}}, \ and\ \bibinfo
  {author} {\bibfnamefont {I.}~\bibnamefont {Franco}},\ }\href {\doibase
  10.1038/s41467-018-04393-4} {\bibfield  {journal} {\bibinfo  {journal}
  {Nature Communications}\ }\textbf {\bibinfo {volume} {9}},\ \bibinfo {pages}
  {2070} (\bibinfo {year} {2018})}\BibitemShut {NoStop}%
\end{thebibliography}%

\end{document}